\documentclass[useAMS,usenatbib]{mn2e}
\usepackage[utf8]{inputenc}
\usepackage[english]{babel}
\usepackage{amsmath}
\usepackage{amssymb}
\usepackage{graphicx}
\usepackage{xcolor}
\usepackage{aas_macros}

\newcommand{\change}[1]{#1}

\newcommand{\href}[2]{#2}

\newcommand{\spotrod}{\texttt{spotrod}}
\newcommand{\macula}{\texttt{macula}}
\newcommand{\prism}{\textsc{prism}}
\newcommand{\python}{\texttt{Python}}
\newcommand{\integratetransit}{\texttt{integratetransit}}
\newcommand{\ellipseangle}{\texttt{ellipseangle}}
\newcommand{\circleangle}{\texttt{circleangle}}
\newcommand{\hatpeleven}{HAT-P-11}
\newcommand{\hatpelevenb}{HAT-P-11b}
\newcommand{\kepler}{\textit{Kepler}}
\newcommand{\zcrit}{\ensuremath{z_\mathrm{crit}}}
\newcommand{\xs}{\ensuremath{x_\mathrm s}}
\newcommand{\ys}{\ensuremath{y_\mathrm s}}
\newcommand{\titlestring}{\change{\spotrod{}:} a semi-analytic model for transits of spotted stars}
\newcommand{\totalspots}{203}

\title[\titlestring]{\titlestring}

\author[Béky et al.]
{Bence Béky$^1$,
David M.~Kipping$^{1,2}$,
Matthew J.~Holman$^1$ \\
$^1$Harvard--Smithsonian Center for Astrophysics, 60 Garden St, Cambridge, MA 02138, USA \\
$^2$Carl Sagan Fellow}

\begin{document}

\pagerange{\pageref{firstpage}--\pageref{lastpage}} \pubyear{2014}

\maketitle

\label{firstpage}

\begin{abstract}
The Hubble Space Telescope (HST) and the \kepler{} space mission observed a large number of planetary transits showing anomalies due to starspot eclipses, with more such observations expected in the near future by the K2 mission and the Transiting Exoplanet Survey Satellite (TESS).  To facilitate analysis of this phenomenon, we present \texttt{spotrod}, a model for planetary transits of stars with an arbitrary limb darkening law and a number of homogeneous, circular spots on their surface.  A free, open source implementation written in \texttt{C}, ready to use in \python{}, is available for download.

We analyze \kepler{} observations of the planetary host star \hatpeleven{}, and study the size and contrast of more than two hundred starspots.  We find that the flux ratio of spots ranges at least from 0.6 to 0.9, corresponding to an effective temperature approximately 100 to 450 K lower than the stellar surface, although it is possible that some spots are darker than 0.5.  The largest detected spots have a radius less than approximately 0.2 stellar radii.
\end{abstract}


\section{Introduction}
\label{sec:introduction}

In a transiting planetary system, spots on the face of the host star can result in deviations in the transit lightcurve from the well-known model described by \citet{2002ApJ...580L.171M}. An unocculted spot --- since it is darker than the stellar surface --- causes a blend in the opposite sense as a background star, leading to a deeper transit \citep[see, for example,][]{2009A&A...505.1277C}. On the other hand, a spot causes an anomalous rebrightening when it is eclipsed by the planet, because the planet blocks less flux than it would if the spot was not behind it.

Such spot-induced transit lightcurve anomalies were first observed by \citet{2003ApJ...585L.147S} in HST observations of HD 209458. Other systems exhibiting similar features include HD 189733 \citep{2007A&A...476.1347P} and TrES-1 \citep{2009A&A...494..391R}. \citet{2009A&A...505.1277C} found a correlation between stellar brightness and transit depth in the system CoRoT-2, which they attribute to varying levels of stellar activity, and show how this effect, when unaccounted for, causes a bias in the planet size estimate.

In the era of the \kepler{} satellite, a large number of planets transiting active stars have been discovered and observed with high temporal and photometric resolution, providing further examples of transit anomalies. Two such systems are Kepler-17 \citep{2011ApJS..197...14D} and \hatpeleven{} \citep{2010ApJ...710.1724B}. Spots revealed by transit anomalies can be used, for example, to constrain the projected obliquity and the stellar inclination \citep{2011ApJ...740...33D,2011ApJ...743...61S}. Measuring the spot contrast allows one to constrain the temperature of the spots \citep{2003ApJ...585L.147S,2009A&A...494..391R}. However, there is a degeneracy between the spot size and contrast \citep{2007A&A...476.1347P,2013MNRAS.428.3671T}, which makes high quality data necessary to infer temperatures. 

The large number of photometric observations of transit anomalies motivates the development of astrophysical models. Examples include the model by \citet{2003ApJ...585L.147S}, the one by \citet{2009A&A...504..561W}, SOAP-T by \citet{2013A&A...549A..35O}, and \prism{} by \citet{2013MNRAS.428.3671T}. These models all assume homogeneous, circular spots, with four input parameters for each spot (two for position, one for size, one for darkness). The first two models simplify the geometry by assuming that the spots are circular in projection, while the other two properly account for the elliptical projected shape given spots that are circular on the stellar surface. These models all define a large resolution two dimensional grid either on the stellar surface or in the projection plane, and numerically integrate over two coordinates to calculate the transit lightcurve.

Integration in two dimensions can be computationally expensive. \citet{2012MNRAS.427.2487K} introduced \macula{}, an analytic model for a related but different phenomenon: spots on the rotating stellar surface modulating out-of-transit lightcurves of spotted stars. Its analytic nature makes \macula{} faster than numerical models for the same phenomenon, like SOAP \citep{2012A&A...545A.109B}.

In this paper, we present \spotrod{}, a counterpart of \macula{} for transit lightcurves of spotted stars. We describe the problem as a two dimensional integral in polar coordinates in the projection plane. Using assumptions similar to those of previous models, we derive an analytic formulation for the integral with respect to the polar angle, so that numerical integration needs to be performed only with respect to the radial coordinate. This semi-analytic nature provides improved speed over previous models requiring two dimensional numerical integration.  In particular, if the resolution of the integration grid is $n$ in each dimension, then a double numerical integral takes $\mathcal O\left(n^2\right)$ time to evaluate, whereas \spotrod{} runs in $\mathcal O(n)$ time.  Typical values are 
$n\approx300$ for SOAP,
$n\approx750$ for the model of \citet{2003ApJ...585L.147S},
and the grid spacing being one hundredth of the planet diameter for \prism{}, resulting in $n\approx1000$ for a typical hot Jupiter or $n\approx2000$ for \hatpelevenb{}.  In this work we use $n=1000$.

We describe the semi-analytic model in Section \ref{sec:geom}: we state the simplifying assumptions, describe the two dimensional integral in polar coordinates, and introduce the subroutines of \spotrod{}, the free and open source implementation available for the astronomical community.  In Section \ref{sec:application}, we apply \spotrod{} to \kepler{} observations of \hatpeleven{}, investigate model artifacts like observational biases and correlations of fit parameters for individual spots, validate \spotrod{} on synthetic data generated by \prism{}, study the distribution of spot size and contrast on \hatpeleven{} that we believe to be physical, and look at the model residuals for validation.  Section \ref{sec:conclusion} concludes our findings.  Technical details of the model pertaining to calculating angles and handling spots that are partially behind the limb are given in Appendix \ref{sec:derivations}.

\section{Spot anomaly model}
\label{sec:geom}

\subsection{Assumptions}
\label{sec:assumptions}

Our model has two major assumptions. The first one is that the boundary of each spot is a circle on the surface of the spherical star. We define the radius $a$ of the spot to be the radius of this circle in three-dimensional space, in units of stellar radius. We assume that $0<a<1$. Note that the analytic rotational modulation model \macula{} \citep{2012MNRAS.427.2487K} takes as input parameter the half angle $\alpha$ of the cone with this circle as its directrix and the center of the star as its apex, which is related to the radius by $a = \sin\alpha$.

We define the center of the spot as the intersection point of the surface of the sphere and the axis of this cone. Note that the center of the spot does not lie in the plane of the boundary. The advantage of this definition is that the center is on the stellar surface, allowing for easier conversion between input parameters of \spotrod{} and \macula{}, and easier treatment of stellar rotation.  A further advantage of characterizing the spot location with the projection of a point on the stellar surface over the projection of the geometrical center of the spot boundary in the interior of the star is that its domain does not depend on the spot radius, which also makes it easier to define an isotropic prior for the location of the spot.

The second assumption is that each spot is homogeneous and observes the same limb darkening law as the star. This means that as viewed by the observer, the ratio of the flux from a spot and the flux from the unspotted stellar surface at the same projected distance $r$ from the center of the star does not depend on the distance $r$. We denote this dimensionless flux ratio by $f$ in accordance with \citet{2012MNRAS.427.2487K}.  Note that flux ratio is sometimes called contrast, for example, by \citet{2013MNRAS.428.3671T}.

Assuming a constant flux ratio across the stellar disk is consistent with the findings of \citet{2003SoPh..213..301W}: they study a sample of 18\,000 spots on the Sun, and observe no dependence of spot contrast on where the spot is seen.  Note that what they call contrast can be expressed as $f-1$ using our notation.  As for the homogeneity of spots, we shall see in Section \ref{sec:penumbra} how to compose more complicated structures, like a spot with umbra and penumbra, using two homogeneous spots.


\subsection{Integration}

Let $I(r)$ denote the stellar intensity according to the limb darkening law up to an arbitrary scaling factor, where $0\leqslant r \leqslant 1$ is the projected distance from the center of the star in units of stellar radius. Let $C_r$ denote the circle of radius $r$ in the projection plane concentric with the stellar disk. Then the total out-of-transit flux of the unspotted stellar surface can be calculated as a two dimensional integral over the polar coordinates $(\vartheta, r)$, with the inner integral along $C_r$, and the outer integral with respect to the radial coordinate:
\begin{align}
\label{eq:nospot}
F_0 &= \int_0^1\int_0^{2\pi} I(r) \mathrm d\vartheta r \mathrm dr = \int_0^1 2\pi I(r) r\mathrm dr.
\end{align}
Here $\mathrm d\vartheta r \mathrm dr$ is the area of an infinitesimal element in the projection plane. The integrand does not depend on $\vartheta$, therefore the inner integral can be evaluated as the product of the integrand $I(r)$ and the length $2\pi$ of the integration interval.

\begin{figure}
\begin{center}
\includegraphics*[width=80mm]{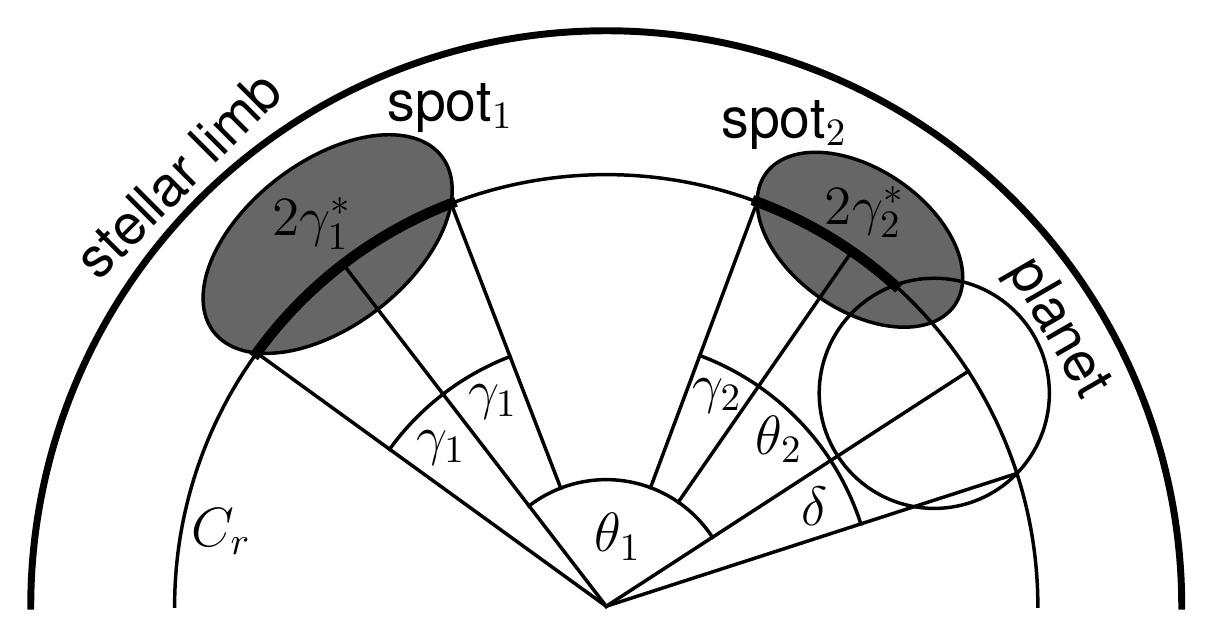}
\end{center}
\caption{Example configuration with transiting planet and two spots, showing $\gamma_i$, $\gamma^*_i$, and $\delta$ for a given value of $r$. $\theta_i$ is used to calculate $\gamma^*_i$, see Appendix \ref{sec:calc}.}
\label{fig:observer}
\end{figure}

Now consider the case of a single spot visible on the star. Let $\gamma(r)$ be half the central angle of the arc of $C_r$ that overlaps with a spot, as shown on Figure \ref{fig:observer}. Then the total flux is
\begin{align}
\nonumber
F_1 &= \int_0^1\left(\int_0^{2\gamma(r)} fI(r)\mathrm d\vartheta + \int_{2\gamma(r)}^{2\pi}I(r)\mathrm d\vartheta\right) r\mathrm dr \\
\nonumber
&= \int_0^1\left(2\gamma(r)fI(r) + (2\pi - 2\gamma(r))I(r)\right)r\mathrm dr \\
\label{eq:factoredout}
&= \int_0^12(\pi + (f-1)\gamma(r))I(r)r\mathrm dr.
\end{align}
Here the inner integral is composed of two parts: inside the spot, on an arc of total length $2\gamma(r)$, the intensity is $fI(r)$, whereas outside the spot, on the remaining arc of length $2\pi-2\gamma(r)$, the intensity is $I(r)$. The integrands do not depend on $\vartheta$, therefore each integral reduces again to the product of the integrand and the length of the corresponding interval. In the final step, we factor out $2I(r)$, and collect the terms with $\gamma(r)$. 

If there are $s$ non-overlapping spots, with corresponding flux ratios $f_i$ and half central angle functions $\gamma_i(r)$, then each inner integral evaluates to $2\gamma_i(r)f_iI(r)$, and the unspotted stellar surface will have an arc length of $2\pi-\sum_{i=1}^s2\gamma_i(r)$, giving a total flux of
\begin{align*}
F_s &= \int_0^1\left(\sum_{i=1}^s2\gamma_i(r)f_iI(r) + \left(2\pi - \sum_{i=1}^s2\gamma_i(r)\right)I(r)\right)r\mathrm dr.
\end{align*}
Just as we factored out $2I(r)$ and collected $\gamma(r)$ in Equation (\ref{eq:factoredout}), we can do the same for each $\gamma_i(r)$ to account for the contribution of multiple spots in a single summation:
\begin{align}
\label{eq:multispot}
F_s &= \int_0^12\left(\pi + \sum_{i=1}^s(f_i-1)\gamma_i(r)\right)I(r)r\mathrm dr.
\end{align}

During transit, let $\delta(r)$ be the half central angle of the arc of $C_r$ that is obscured by the planet. Let $\gamma^*_i(r)$ be the half central angle of the arc on the same circle that overlaps with spot $i$, but is not obscured by the planet. See spot 2 on Figure \ref{fig:observer} for an example. Then the total flux can be calculated by substituting $\gamma^*_i(r)$ for $\gamma_i(r)$ in Equation (\ref{eq:multispot}), and subtracting $2\delta(r)$ from the arc length of the unspotted stellar surface:
\begin{align}
\nonumber
F_\mathrm{transit} &= \int_0^1\left(\sum_{i=1}^s2\gamma^*_i(r)f_iI(r) + \right. \\
\nonumber
&\phantom{=} \quad + \left.\left(2\pi - \sum_{i=1}^s2\gamma^*_i(r) - \change{2}\delta(r)\right)I(r)\right)r\mathrm dr \\
\label{eq:transit}
&= \int_0^12\left(\pi - \delta(r) + \sum_{i=1}^s(f_i-1)\gamma^*_i(r)\right)I(r)r\mathrm dr.
\end{align}
The contribution of the planet in this formula is formally equivalent to a spot with flux ratio $f=0$. 

The final product of our proposed model is the dimensionless normalized transit lightcurve
\begin{align}
\label{eq:normalized}
F_\mathrm{normalized} &= \frac{F_\mathrm{transit}}{F_s}.
\end{align}
At this step, the arbitrary scaling factor in $I(r)$ cancels out.

Note that we define $\gamma_i(r)$ and $\gamma^*_i(r)$ to be zero in case the corresponding arcs do not \change{exist}, that is, $C_r$ does not intersect the spot in projection. Similarly, $\delta(r)$ is understood to be zero if the planet does not eclipse $C_r$. See Appendix \ref{sec:calc} on calculating $\gamma_i$, $\gamma^*_i$, and $\delta$.

In case there are no spots on the stellar surface, even though our model still yields the correct lightcurve asympotically for large grid resolution $n$, we suggest using the fully analytic algorithm of \citet{2002ApJ...580L.171M} if speed is a consideration.

\subsection{Implementation}

The model described in this paper is implemented as a software package called \spotrod{}. It provides four functions:
\begin{itemize}
\item[] \texttt{elements} takes the planetary period, semi-major axis, $k=e\cos\varpi$, $h=e\sin\varpi$, and an array of observation times of a transiting planet with respect to the time of midtransit as input parameters, and calculates the arrays of planar orbital elements $\xi$ and $\eta$ using the formalism of \citet{2009MNRAS.396.1737P}.
\item[] \circleangle{} takes the planetary radius $R_\mathrm p$, the distance $z$ of the centers of the planet and the stellar disk in projection plane, and an array of radii $r$ as input parameters, and calculates the array $\delta(r)$.
\item[] \ellipseangle{} takes the projected spot semi-major axis $a$, the distance $z$ of the centers of the projected spot boundary and the stellar disk in projection plane, and an array of $r$ as input parameters, and calculates the array $\gamma(r)$. This function is executed internally by \integratetransit{}, so the user does not have to call it direcly.
\item[] \integratetransit{} takes $R_\mathrm p$, arrays of the projected coordinates of the planet and the spots, spot radii and flux ratios, an array of the radii $r$ and weights for numerical integration (the latter are calculated from the integration quadrature and the limb darkening law), and precalculated values of $\delta(r)$ for each value of $r$ and each observation time as input parameters. It calculates the normalized lightcurve $F_\mathrm{normalized}$, using analytic integration with respect to the polar angle, and numerical integration with respect to $r$.
\end{itemize}

The software can employ any integration quadrature, that is, numerical method that works by evaluating the integrand at given values of $r$ and summing up using given weights.  It evaluates the inner integral analytically in the form of the sum given in Equations (\ref{eq:multispot}) and (\ref{eq:transit}) on each annulus defined by the input array of $r$ values.  Then it performs numerical integration with respect to $r$ by adding up the products of these values and the corresponding weights.

Among the simplest integration quadratures are the trapezoidal rule and the midpoint rule.  We recommend an integration mesh of $n\approx1000$ values uniformly spaced between 0 and 1.  More complicated rules can also be prescribed.  For example, since the integrand of the outer integral grows rapidly with $r$ in Equations (\ref{eq:nospot}-\ref{eq:transit}), one might wish to use a nonuniform mesh that is coarser for small $r$ and finer for large $r$.

\spotrod{} can also handle arbitrary limb darkening laws, even ones without an analytic formula: it only relies on limb darkening values evaluated at the values of $r$ used in the integration rule. The limb darkening law and the integration quadrature weights are then multiplied together before they are passed to \integratetransit{}, since it is only their product that is ever used.

Repeated evaluations of \integratetransit{} at the same observation times with fixed planetary orbital parameters are required, for example, for fitting or a Monte Carlo Markov Chain (MCMC) exploring spot parameters. The code has been optimized for such use: one needs to calculate the arrays $\xi$ and $\eta$, then the projected planetary coordinates at each observation, and finally an array of $\delta$ only once at the beginning. These values do not depend on spot parameters, and recalculating the same values of $\delta$ in each iteration would be very costly, since it has to be calculated for each observation time and each $r$: hundreds of thousands of times in a typical application. Instead, we evaluate $\delta$ hundreds of thousands of times only once, before we start the fit or MCMC, and then use these precalculated values.

We can also avoid evaluating the function \ellipseangle{} hundreds of thousands of \change{times} in each iteration if we neglect the effect of stellar rotation during a single transit. In this case, for a given set of spots, $\gamma_i(r)$ does not depend on time, therefore we only need to calculate it $n$ times: once for each value of $r$.

It is, of course, also possible to model a transit in the extreme case of a star that rotates so rapidly that spots move substantially during the duration of the planetary transit. In this case, one needs to recalculate the spot positions and call the function \integratetransit{} with a time array of length one for each observation. This method, however, is slower than if we assumed that spots were stationary within the duration of a single transit.

\spotrod{} is free and open source software, released under the GNU General Public License. It is implemented in \texttt{C} in the interest of speed, and provides bindings for use in \python{}. Bindings for different programming languages should be reasonably easy to add.

\spotrod{} is publicly available for download at \href{https://github.com/bencebeky/spotrod}{\texttt{https://github.com/bencebeky/spotrod}}, including the \texttt{C} code, \python{} bindings, two example programs in \python{}, compilation instructions, and a copy of the license.

\subsection{Umbra, penumbra, and faculae}
\label{sec:penumbra}

Note that in Equations (\ref{eq:multispot}--\ref{eq:transit}), the contribution of spots add up, regardless of whether they overlap or not. As \citet{2012MNRAS.427.2487K} points out in his Section 2.4, this feature can be used to build a composite spot with a central umbra of radius $a_\mathrm u$ and flux ratio $f_\mathrm u$ and a surrounding penumbra of radius $a_\mathrm p$ and flux ratio $f_\mathrm p$ by feeding two concentric spots with radius-flux ratio pairs $(a_\mathrm p, f_\mathrm p)$ and $(a_\mathrm u, 1 - f_\mathrm p + f_\mathrm u)$ into the model.

Spots should have flux ratio between 0 and 1, $f=0$ for a completely dark spot, and $f=1$ for one indistinguishable from the stellar surface. We note that as the model can handle any value of flux ratio $f$, faculae and plages (bright areas on the stellar photosphere and chromosphere, respectively) can also be modelled using a flux ratio value exceeding 1, as suggested, for example, by \citet{2012A&A...545A.109B, 2012MNRAS.427.2487K}.

\section{Application to \hatpeleven}
\label{sec:application}

\hatpeleven{} is 9.6 visual magnitude K4 dwarf star in the field of the \kepler{} space telescope \citep{2010Sci...327..977B}. It has been known to host \hatpelevenb{}, a transiting hot Neptune on a 4.9 day orbit \citep{2010ApJ...710.1724B}, before the launch of the \kepler{} mission, and therefore has been observed with short (one minute) cadence from the beginning, during quarters 0--6, 9--10, 12--14, and 16--17. Missing data in quarters 7, 11, and 15 are due to the failure of a readout module in 2010 January. \change{In quarter 8, only long cadence observations were taken.}  We perform our analysis on this dataset, using the flux values in the \texttt{SAP\_FLUX} column, and dividing each transit by a linear fit to the out-of-transit data within 0.12 days from the midtransit time to normalize the transit lightcurves.  Visual inspection shows that a linear fit is satisfactory, because the out-of-transit lightcurve modulation timescale is the rotation period of \hatpeleven, which is 29.2 days \citep{2010ApJ...710.1724B,2014ApJ...788.....1}, much larger than the 0.24 days of the total width of our window.

For our analysis, we adopt the orbital eccentricity and argument of periastron values reported by \citet{2010ApJ...710.1724B} based on RV data and Hipparcos parallax for \hatpeleven{}. However, we use the revised transit ephemeris, planetary radius and orbital semi-major axis relative to the stellar radius, orbital inclination, and limb darkening parameters of \citet{2011ApJ...740...33D}.  Their treatment relies on the above eccentricity and argument of periastron values, but accounts for eclipsed and uneclipsed spots, that biased earlier analyses.  We number the transits according to this ephemeris, with the midtransit time of transit 0 being $T_0 = 2\,454\,605.891\,55 \pm 0.000\,13$ (barycentric dynamical time).

Rebrightening events in the transit lightcurve of \hatpeleven{} due to spots were first predicted by \citet{2010ApJ...723L.223W}, and first reported independently by \citet{2011ApJ...743...61S} and \citet{2011ApJ...740...33D}.  They are used to constrain the stellar rotational period by \citet{2014ApJ...788.....1}, who also compare a model out-of-transit lightcurve based on the MCMC chains described in this section, and find that it is consistent with the assumption that out-of-transit variation is dominated by rotation of spots.

\subsection{Analysis of individual spots}
\label{sec:individual}

First, we present the analysis of two individual starspots in order to study correlations between spot parameters inherent to the model. Each spot is described by four parameters: $x$ and $y$ are the coordinates of the spot center in stellar radius units, as seen by the observer, in a Cartesian coordinate system whose origin is the center of the stellar disk, and where the planet is moving approximately in the positive $x$ direction during transit. (More precisely, for inclined orbits the $y$ axis is defined by projecting the line of sight on the orbital plane, then projecting that on the sky plane. For an inclined eccentric orbit, the projected velocity of the planet is not exactly parallel to the $x$ axis at mid-transit, except for special values of the argument of periastron.) The other two parameters are the spot radius $a$ and the flux ratio $f$ described in Section \ref{sec:assumptions}.

\begin{figure}
\begin{center}
\includegraphics*[width=80mm]{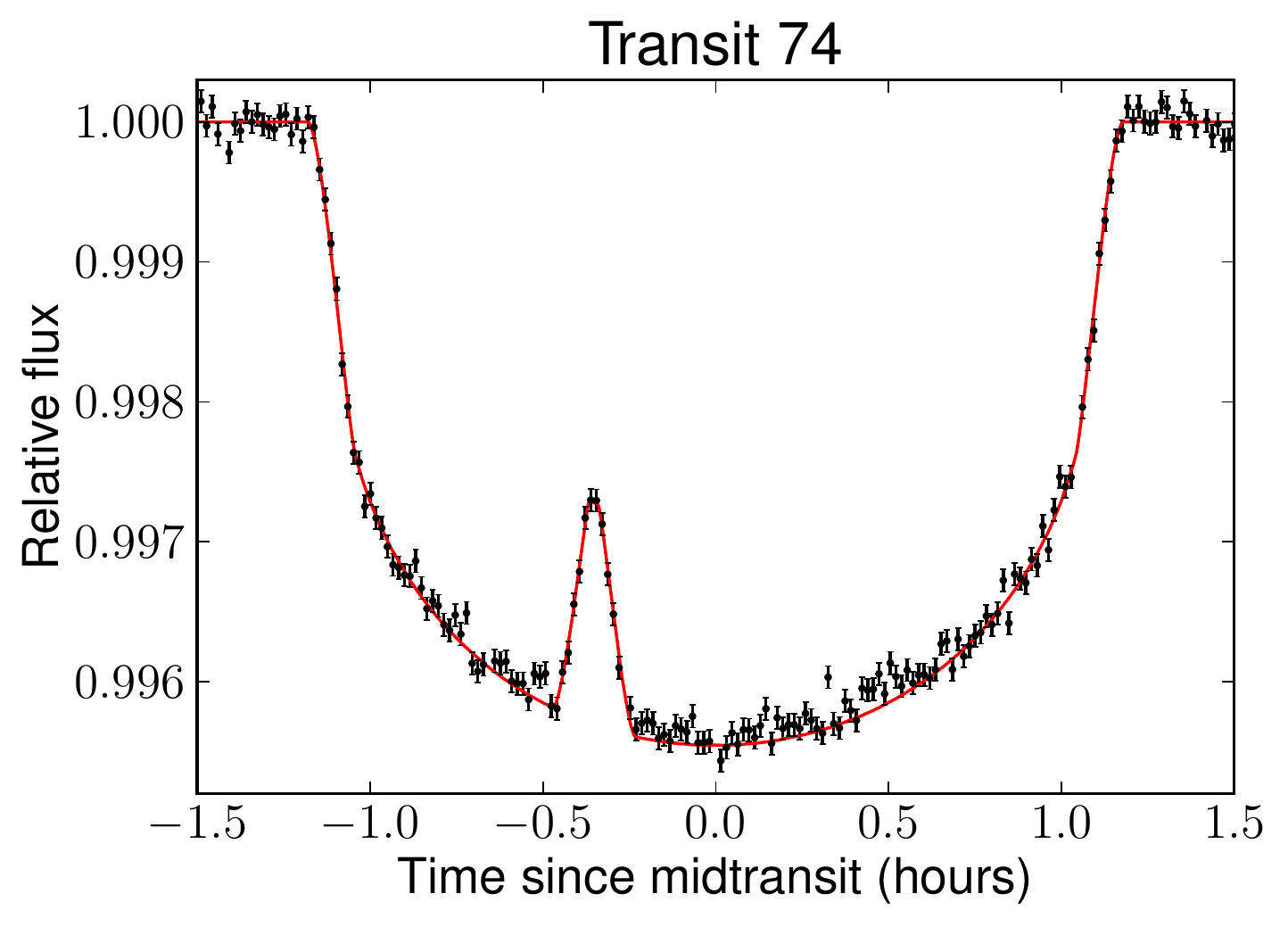}
\includegraphics*[width=80mm]{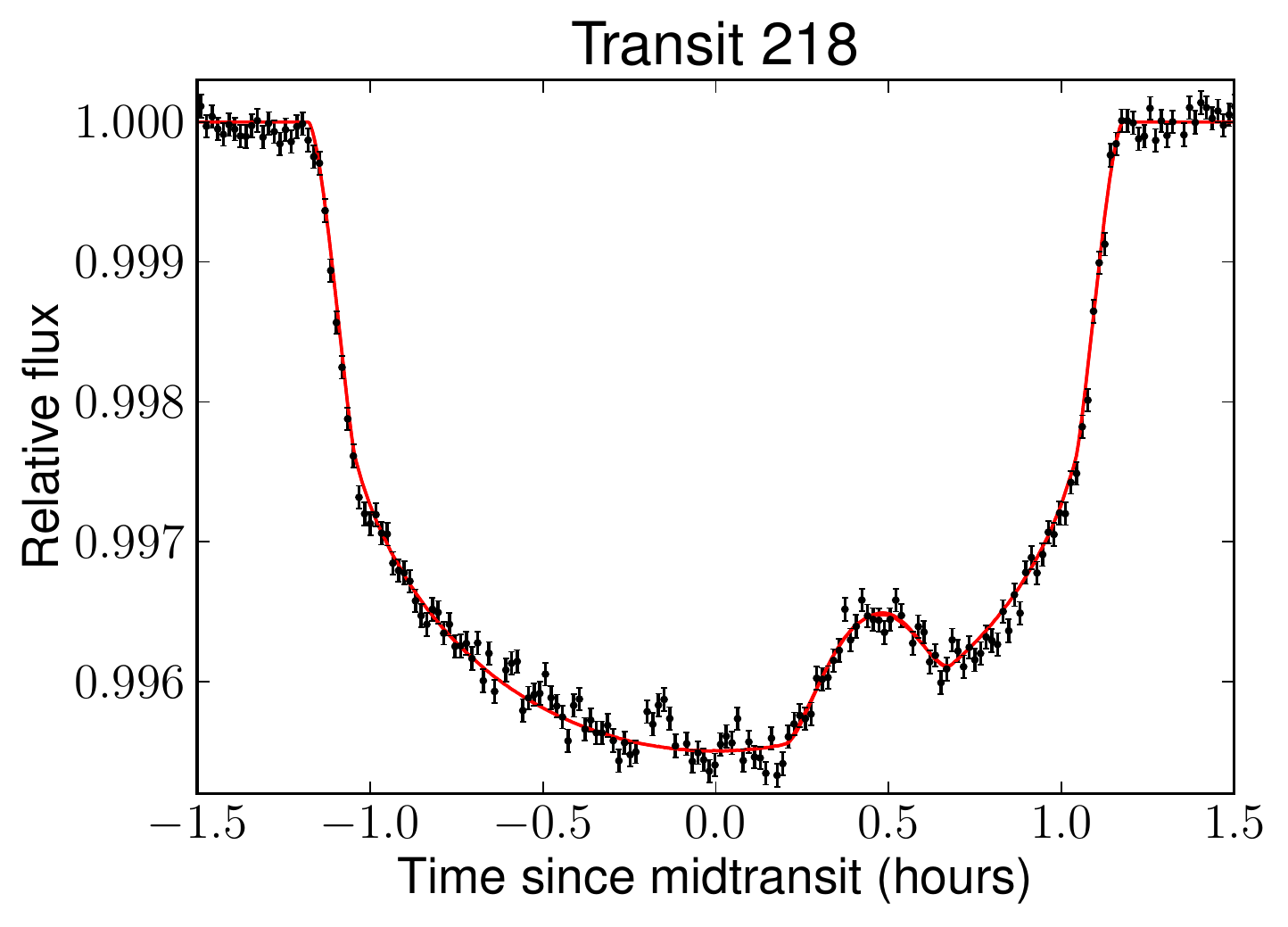}
\end{center}
\caption{\hatpeleven{} transit 74 (top panel) and transit 218 (bottom panel) lightcurves. Dots are \kepler{} short cadence observations, with errorbars given by the \texttt{SAP\_FLUX\_ERR} data column. Red curves are best fit \spotrod{} models assuming a single spot on the stellar surface for both transits.}
\label{fig:transit218}
\end{figure}

Figure \ref{fig:transit218} presents the lightcurves of transits 74 and 218.  For both transits, we identify the deviation from the lightcurve model of \citet{2002ApJ...580L.171M} as an indication for the planet eclipsing a single spot on the surface of the star.  The lightcurve anomalies are observed about $0.4$ hours before midtransit during transit 74, and about half an hour after midtransit during transit 218.  We also plot the best fit (least sum of squared residuals) \spotrod{} lightcurves for both transits in red on this figure.

\begin{figure}
\begin{center}
\includegraphics*[width=80mm]{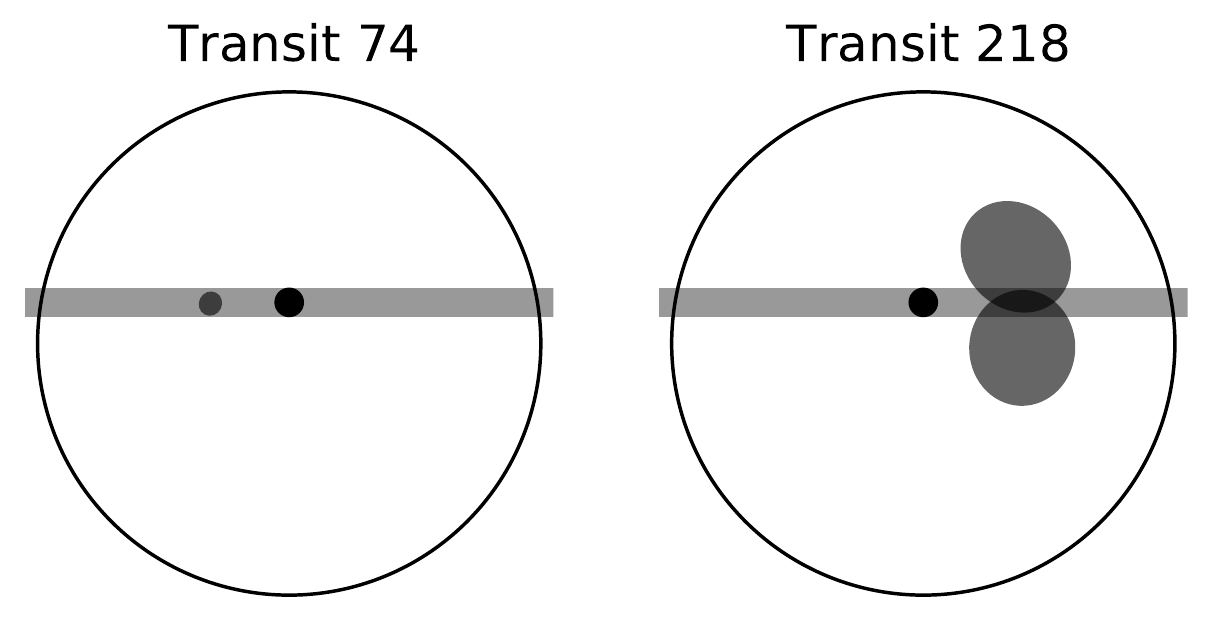}
\end{center}
\caption{Projected images of \hatpeleven{} during transits 74 (left panel) and 218 (right panel). Large circle is the star, solid black small circle is the planet \hatpelevenb{} at midtransit, gray strip is the transit chord, one or two gray ellipses are best fit solutions for the one or two modes of the spot.}
\label{fig:spots}
\end{figure}

In fact, for transit 218, we find two best fit solutions: in the projection plane, the spot can either be situated above or below the transit chord.  To illustrate this bimodality, we present on Figure \ref{fig:spots} an observer's view of the star (large empty circle), the planet (black filled circle), and the best fit solutions for the single spot (gray ellipses) during transits 74 and 218.  The transit chord is also drawn, as wide as the diameter of the planet.  We do not find the same bimodality in case of transit 74, therefore only one solution is depicted.  Note that in fact we plot two model lightcurves for transit 218 on Figure \ref{fig:transit218}, corresponding to the best fits for each mode.  However, they are indistinguisable on this figure, which is closely related to our inability to infer which solution describes the spot in reality.

We run parallel tempered MCMC simulations \citep{2005PCCP....7.3910E} for both transits, using the \texttt{emcee} software package \citep{2013PASP..125..306F}, at 10 different temperatures, with 100 concurrent chains at each temperature. We first run both simulations for 1000 steps for burn-in.  By inspecting the evolution of the mean and scatter of each parameter, we find that the chain already converges after about half this many steps.  Then we run the simulation for another 1000 steps to sample what we believe to be the equilibrium distribution.  The lowest, zero temperature chain provides us with the equilibrium distribution, whereas the higher temperature chains guarantee that we explore the entire parameter space, and have samples in disconnected modes in numbers proportional to the corresponding posterior probabilities.

\begin{figure*}
\begin{center}
\includegraphics*[width=170mm]{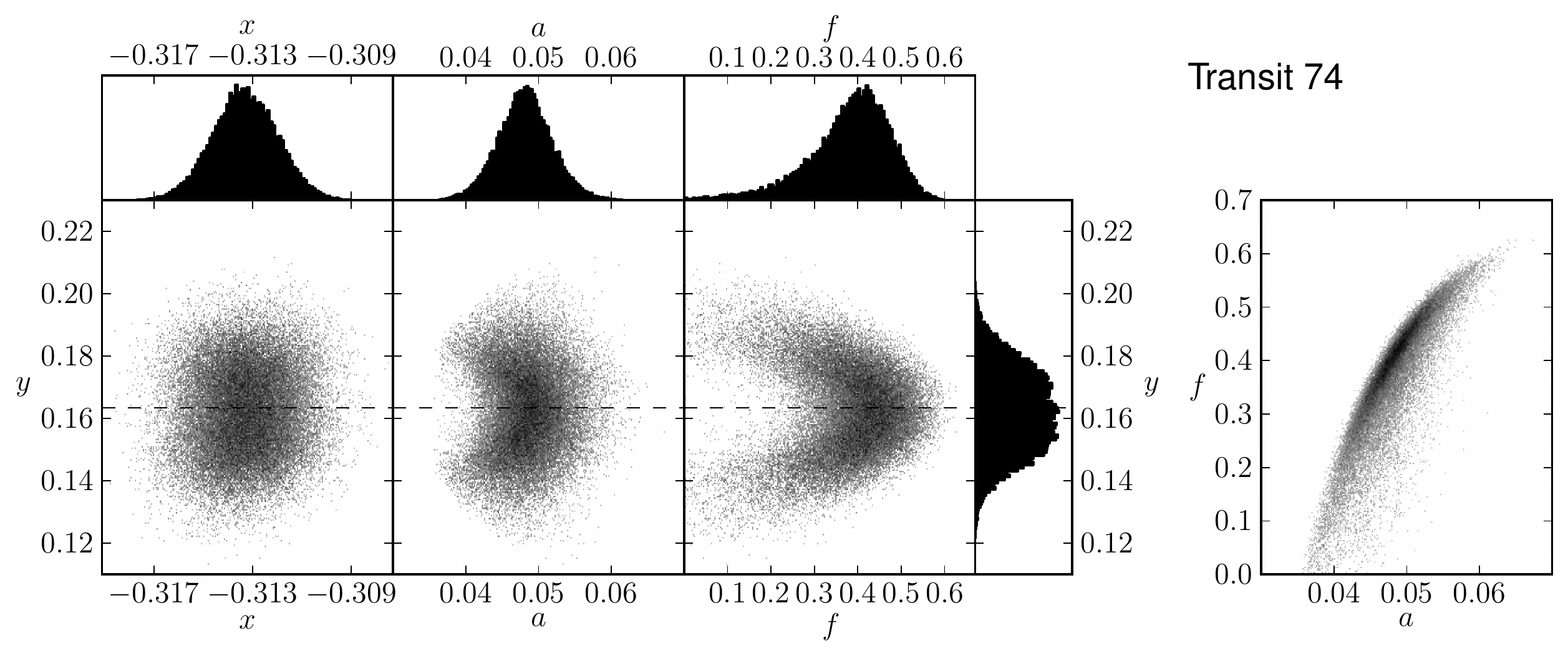}
\includegraphics*[width=170mm]{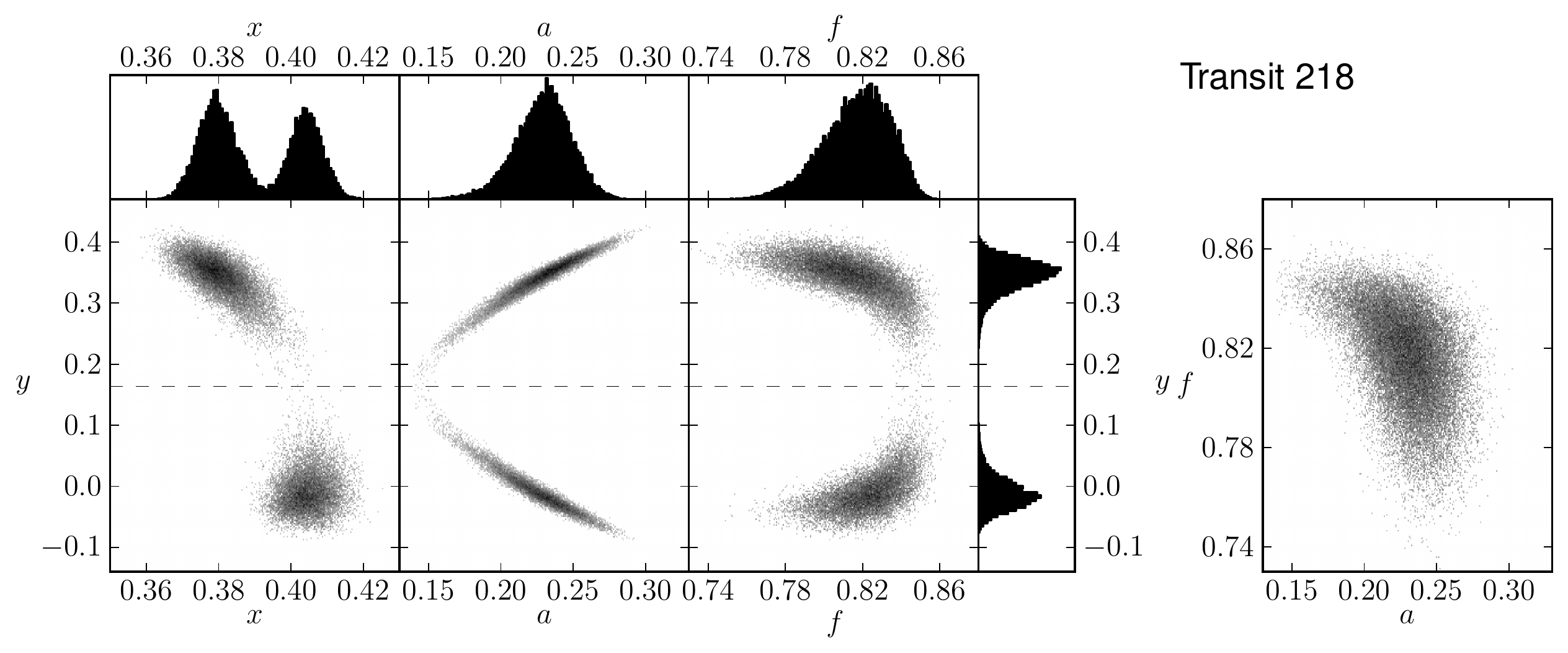}
\end{center}
\caption{Spot parameter distributions of 100\,000 MCMC samples for transits 74 (top panels) and 218 (bottom panels) of \hatpelevenb{}.  Four \change{scatter plots} show joint distributions of pairs of parameters for each transit: $x$--$y$, $a$--$y$, $f$--$y$, and $a$--$f$.  Thin dashed line indicates $y$ coordinate of planet at midtransit.  The four side panels on top and right present histograms of $x$, $a$, $f$, and $y$.}
\label{fig:mcmc}
\end{figure*}

Figure \ref{fig:mcmc} illustrates the results of the MCMC simulation for the single spot models for transits 74 and 218: joint distributions for four pairs of parameters as well as histograms for each parameter are presented.  A dashed line is drawn at the $y$ coordinate of the planet at midtransit, which is very close the the impact parameter $b$ (in fact they would be the same for a circular orbit), to help distinguish the two modes and inspect symmetry with respect to the transit chord.  We immediately confirm that there is only one mode for transit 74, and two modes for transit 218.  The reason for this is that the anomaly during transit 74 can be well described by a spot that lies under the transit chord, therefore the two modes overlap.  On the other hand, the lightcurve of transit 218 can only be well modelled if the spot is further away from the transit chord, in which case the two modes are disconnected.  We note that we experience bimodality for roughly one quarter of all spots in the entire \kepler{} dataset for \hatpeleven{}.

Figure \ref{fig:mcmc} also tells us about the correlations of spot parameters.  First, we note that in both cases, $x$ is better constrained than $y$, even within a single mode.  Recall that $x$ is the coordinate (almost exactly) parallel to the transit chord, therefore it directly relates to when the anomaly is observed, which is well defined by the observations.  On the other hand, we shall see that $y$ correlates with other parameters that together shape the transit anomaly, resulting in a larger uncertainty.

We also note that the two solutions of transit 218 have slightly different best fit $x$ values, even though $x$ is fairly well constrained.  The explanation for this is that since the spot is elliptical in projection, with its semi-major axis not quite parallel to the $y$ axis, therefore the two spot solutions must have different $x$ coordinates in order to intersect the transit chord at roughly the same $x$ coordinate.  This effect can also be observed on Figure \ref{fig:spots}.

The joint distribution of $a$ and $y$ for the spot in transit 218 on Figure \ref{fig:mcmc} tells us that the larger the spot is, the further it has to be from the transit chord.  This is expected from geometrical arguments, as the duration of the transit anomaly can be well constrained from the observations.  This correlation between spot parameters has been first pointed out by \citet{2003ApJ...585L.147S}.  However, this effect does not show for transit 74, probably due to the spot being close to the transit chord.

If we inspect the joint distribution of $f$ and $y$ for either transit, we notice that if the spot is brighter (has a larger flux ratio), then it is likely to be closer to the transit chord.  Such a correlation was first noted by \citet{2009A&A...504..561W}, and it remains to be explained.

Finally, we notice that there is a strong correlation between $a$ and $f$ for transit 74: the brighter the spot is, the larger it has to be.  This correlation has been reported by \citet{2007A&A...476.1347P}, \citet{2009A&A...504..561W} and \citet{2013MNRAS.428.3671T}.  This phenomenon is naïvely explained by that the rebrightening amplitude is proportional to how much the flux blocked by the planet is less than it would be for the unspotted photosphere.  If a spot is not that much darker than the typical stellar surface, a larger area is required to produce the same flux deficit.  This argument is expected to hold for spots centered on the transit chord, for which the spot radius directly determines the occulted area.  On the other hand, we do not observe the same $a$--$f$ correlation for transit 218, becase $a$ correlates strongly with $y$, resulting in a more complex effect on the occulted spot area.

\subsection{Test on synthetic lightcurves}
\label{sec:synthetic}

In this section, we generate synthetic transit lightcurves, and apply to them the same analysis as in the last section.  We use \prism{} \citep{2013MNRAS.428.3671T} for generating lightcurves, and analyze them using \spotrod{}.  Using different models for data generation and analysis allows us to validate them \change{against} each other, that is, make sure that parameters like spot radius and flux ratio are interpreted identically, and they produce the same result.

We take the best fit parameters of the spots in transits 74 and 218, convert the projected coordinates to equatorial coordinates as expected by \prism{}, and generate two model transits.  The difference from the \spotrod{} model with the same input parameters has mean $10^{-8}$ and standard deviation $2\cdot10^{-6}$ across all observation times, which indicates a good agreement between the two models.  For comparison, the mean photon noise is $8\cdot10^{-5}$ for the same data points.

\begin{figure*}
\begin{center}
\includegraphics*[width=170mm]{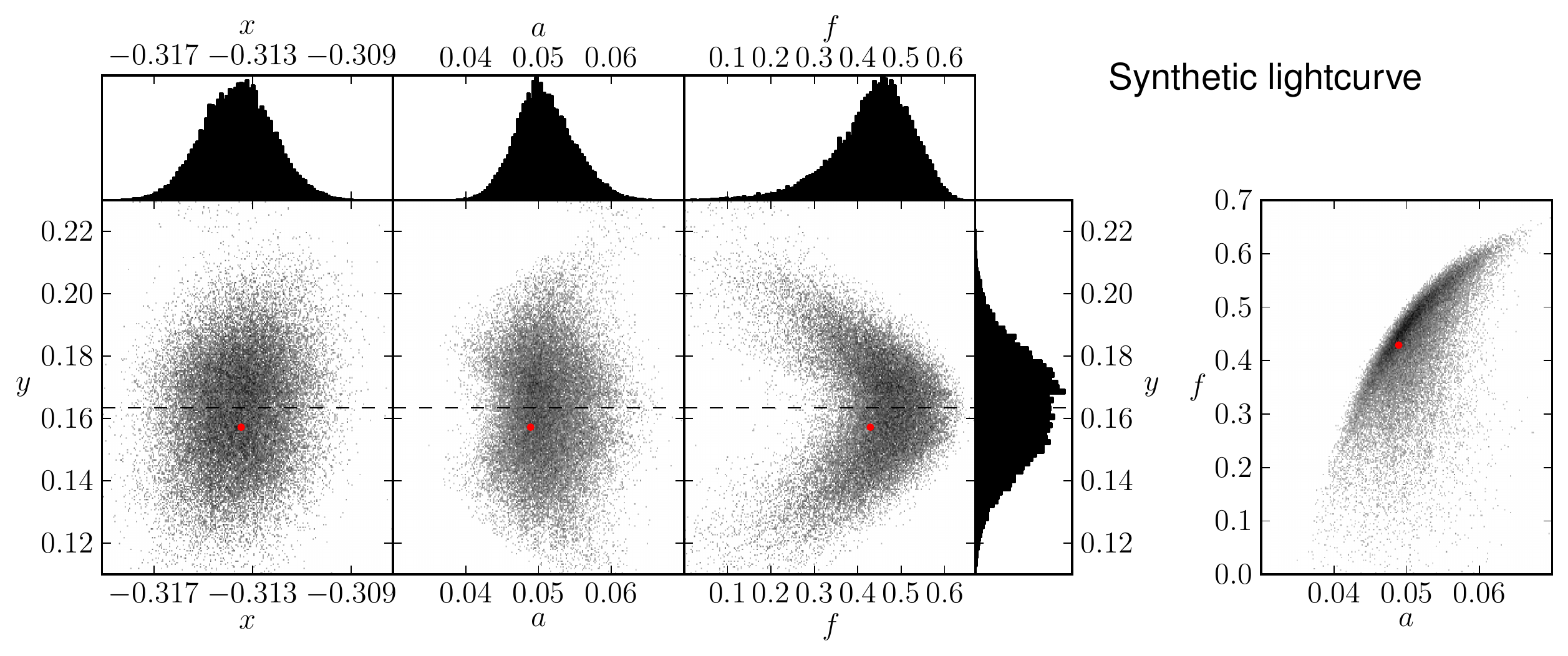}
\includegraphics*[width=170mm]{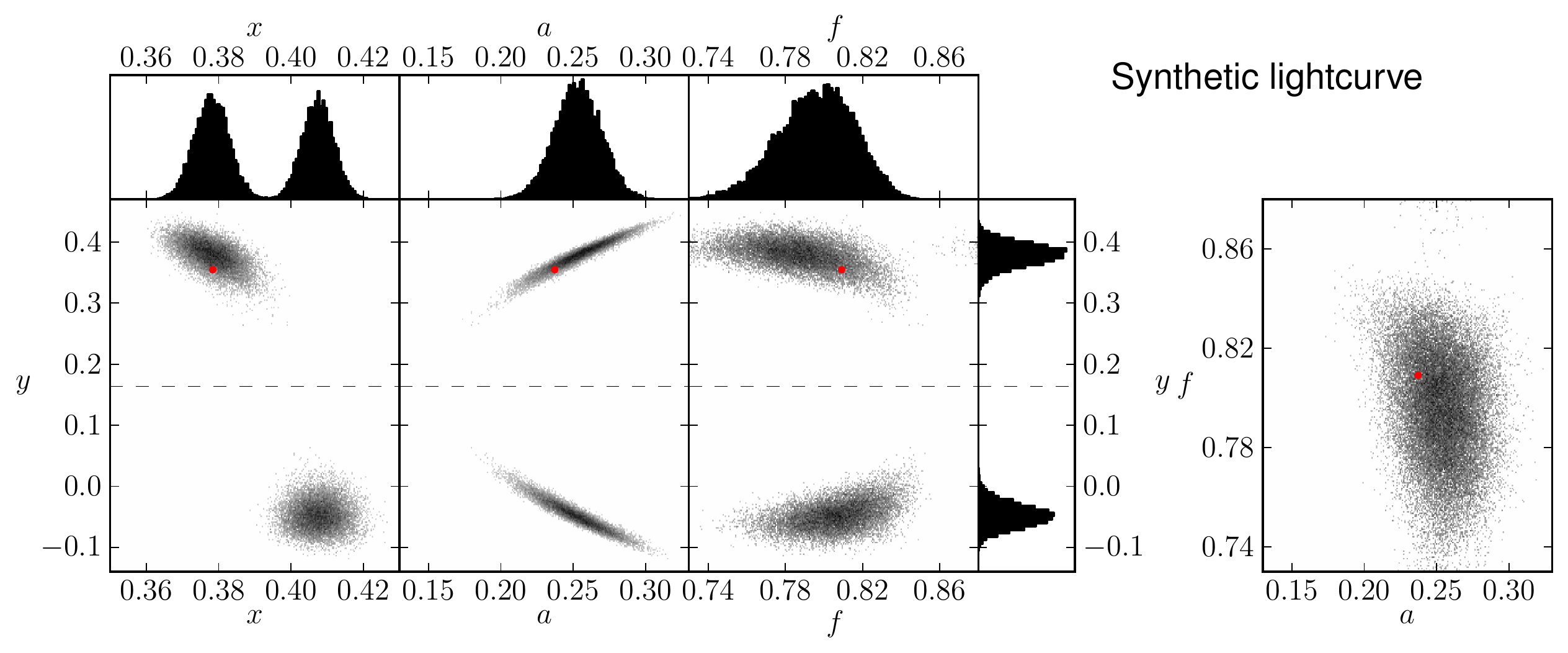}
\end{center}
\caption{Same as Figure \ref{fig:mcmc}, for synthetic transit lightcurves generated using \prism{}, with the best fit spot parameters of the spot from transit 74 (top panels) and 218 (bottom panels) of \hatpelevenb{}.  Red points indicate the best fit spot parameters.}
\label{fig:synthetic}
\end{figure*}

We find that the \spotrod{} best fit has residuals with a standard deviation approximately 1.3 times that predicted by the \texttt{SAP\_FLUX\_ERR} data column of the \kepler{} dataset, therefore we add independent, normally distributed noise scaled to 1.3 times the corresponding \texttt{SAP\_FLUX\_ERR} value to each data point calculated by \prism{}.  We run an MCMC simulation on the resulting synthetic lightcurves for both transits, in a fashion identical to that explained in the previous section.  The resulting chain distribution is presented on Figure \ref{fig:synthetic}.

Comparing Figures \ref{fig:mcmc} and \ref{fig:synthetic}, we find that the chains converge to roughly the same parameters, further confirming that \prism{} and \spotrod{} interpret input parameters in compatible ways.  We also find that the extent and shape of the equilibrium distributions, that is, the correlations between spot parameters, are roughly the same.  In case of transit 218, even though the synthetic lightcurve is generated based on only one of the two modes, we are unable to determine which mode it is from the MCMC analysis: the distribution is bimodal, just like it was for \kepler{} observations.

The ultimate test to decide whether bimodality and parameter correlations are inherent properties of lightcurve models, and not unique to the implementation we use, would be to run MCMC analysis using \prism{} or another model, other than \spotrod{}.  However, since all previously known models require numerical integration in two dimensions, this would be prohibitively computationally expensive.  Instead, we rely on the very small difference of the two lightcurve models when run with the same input parameters to conclude that \spotrod{} reproduces the results of \prism{}.

\subsection{Distribution of spot parameters}
\label{sec:spotparameterdistribution}

In order to study the spot ensemble distribution, we look for anomalies in 204 transits of \hatpelevenb{} of which there are complete, high quality short cadence \kepler{} data.  First, we identify transit anomalies by visual inspection, fit spot parameters using guesses as initial values, and run MCMC simulations.  In a few cases, we find that the chain abandons our initial fit and converges to a solution of a much larger spot with flux ratio close to one, representing a much longer duration and smaller amplitude lightcurve anomaly.  We attribute this to either noise or the effect of a large number of small spots, neither of which we prefer to incorrectly treat as a single, very large spot.  We therefore decrease the number of modelled spots by one for such transits, or omit transits that exhibit such a behaviour with a single spot model, in five cases in total.  For another ten transits, at most 25\% samples of the chain are in an isolated mode representing such an unphysical spot, which samples we discard while keeping the rest of the chain.

We end up with \totalspots{} spots in 130 transits.  For each transit, we independently run the same parallel tempered MCMC simulations as described in Section \ref{sec:individual}, using the \texttt{emcee} software package, at 10 different temperatures.  For 73 transits with one spot each, we use 100 concurrent chains, for 43 transits with two spots each, 200 concurrent chains, for 12 transits with three spots each, we employ 300 chains, and for 2 transits with four spots each, 400 chains.  We have four dimensions of parameter space for each spot: $x$, $y$, $a$, and $f$.  We run the simulations for 1000 steps that we discard.  Again, inspection of the chain shows that roughly half of this is already enough for convergence.  Then we run the chain for another 1000 steps to sample the supposedly equilibrium distribution.  This yields the dataset that we use for our analysis in this paper, and the same dataset is used by \citet{2014ApJ...788.....1}.

The next step is to quantify how much better fit these models provide than if we modelled the lightcurve without the spots.  In order to do so, we calculate the Bayesian Information Criterion (BIC), which is the sum of $\chi^2$ and an additional term penalizing extra model parameters to avoid overfitting \citep {Schwarz1978}.  We find that every single one of the \totalspots{} spots yields a BIC value at least $25.0$ lower than the model without that spot, which is a very significant improvement, justifying every spot in our analysis.  This suggests that probably more spots could be carefully included.

\begin{figure}
\begin{center}
\includegraphics*[width=80mm]{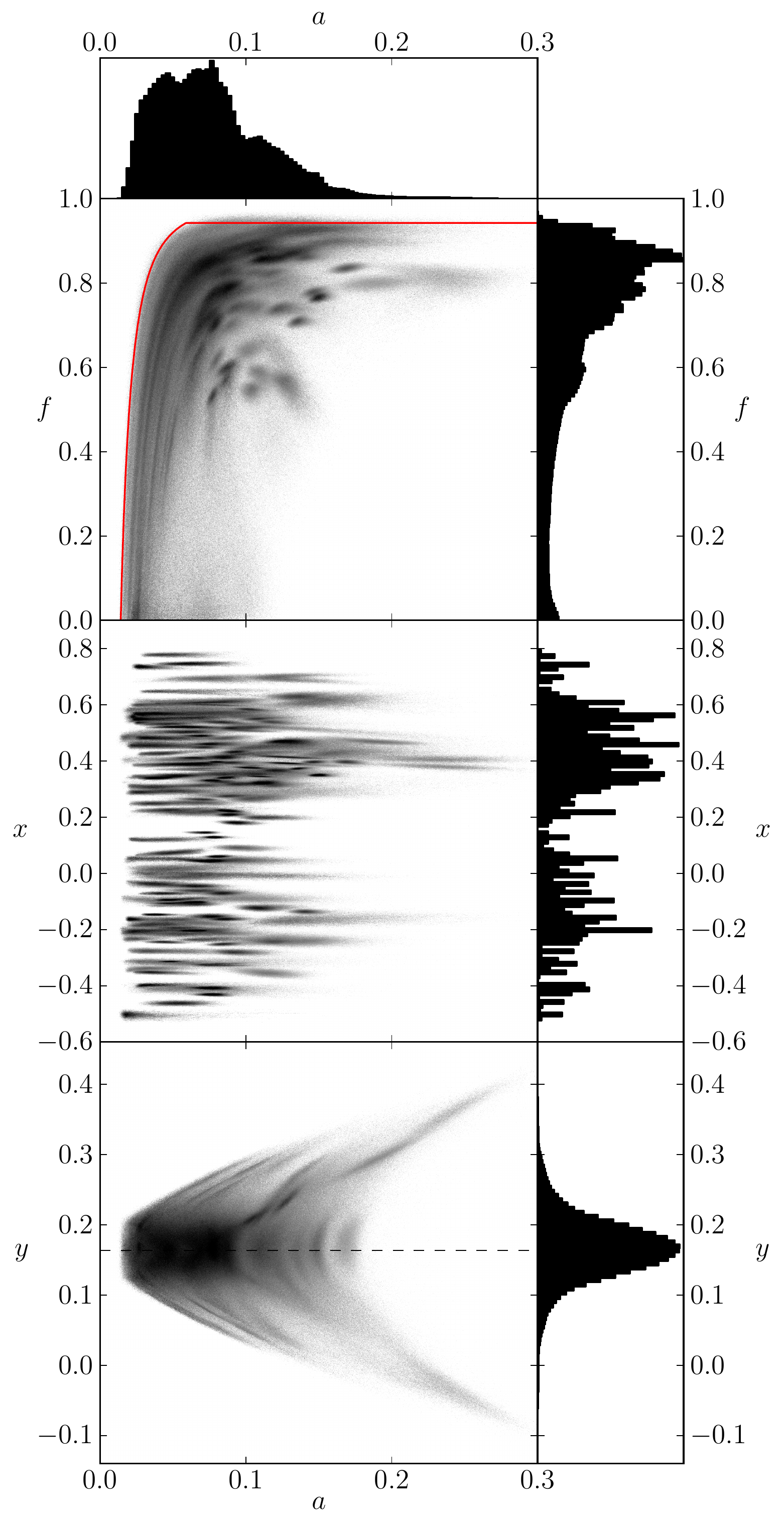}
\end{center}
\caption{\change{Spot parameter distributions of the entire MCMC chain of all \totalspots{} spots.  Three scatter plots present joint distributions of $a$--$f$, $a$--$x$, and $a$--$y$.  On the top scatter plot, red curve is the transit anomaly amplitude threshold $A_\mathrm{max}=0.0002$.  On the bottom scatter plot, thin dashed line indicates $y$ coordinate of planet at midtransit.  The four side panels on top and right present histograms of $a$, $f$, $x$, and $y$.}}
\label{fig:mcmcall}
\end{figure}

We present the spot radius--flux ratio distribution of all spots on Figure \ref{fig:mcmcall}.  Note that the distribution of most individual spots overlap, except for a few, which appear as separate clusters on the figure.  The distribution is bounded from the side of small radius and large flux ratio by an observational bias: whether we visually inspect the lightcurves or apply an algorithmic transit anomaly search, small amplitude anomalies will be lost in photon noise.  The maximum eclipsed spot area is the smaller of $R_\mathrm p^2$ and $a^2$ relative to the stellar disk, therefore the maximum transit anomaly amplitude for given $a$ and $f$ is $A_\mathrm{max} = \min\left(R_\mathrm p^2, a^2\right) (1-f)$ (not accounting for limb darkening and projection distortion).  The actual amplitude of the transit anomaly can be less, depending on $y$.  Figure \ref{fig:mcmcall} shows the $A_\mathrm{max}=0.0002$ curve in red, where $0.0002$ is an arbitrarily chosen value, which seems to act as an approximate amplitude threshold for anomalies that we detected.

We note that for small spots, individual spot distributions spread along constant $\min\left(R_\mathrm p^2, a^2\right)(1-f)$ curves.  The reason is that this quanitity describes the total amount of flux missing due to the spot, which is directly related to the shape of the transit lightcurve anomaly, and therefore can be well constrained.  For small spots, the anomaly does not last long, therefore there are fewer data points than for large spots.  This makes it difficult to resolve the degeneracy between radius and flux ratio.  For large spots, however, the transit anomaly has a larger amplitude, which can directly be used to infer the flux ratio.  Note that consistently with this argument, the rightmost panels of Figure \ref{fig:mcmc} show that the flux ratio of the small spot seen in transit 74 has a much larger uncertainty than that of the large spot seen in transit 218, even though they have comparable relative uncertainties of their radii.

This effect results in small spots having a weaker constraint on flux ratio.  However, as we will see later in this section, this does not indicate the presence of small, dark spots.  Because of this large flux ratio uncertainty of small spots, we are not able to detect a significant correlation between spot radius and flux ratio for the \totalspots{} spots studied.

\change{Figure \ref{fig:mcmcall} also presents the spot radius--$x$ coordinate joint distribution of all identified spots.  We find two active latitudes, at $x\approx-0.2$ and $0.4$, where there are the most spots, and where spots seem to be the largest.  This is in agreement with the findings of \citet{2011ApJ...740...33D,2011ApJ...743...61S}. Note how this panel illustrates that the $x$ coordinate can be very well constrained by the time of transit anomalies.}

\change{Finally,} the bottom left panel of Figure \ref{fig:mcmcall} shows the joint distribution of spot radius and $y$ coordinate for all spots.  Because of the polar orbit of \hatpelevenb{}, the latter roughly corresponds to the longitude of the spot on the stellar surface.  In this case, we can assume that there is no physical correlation between the two parameters, therefore we have to interpret the joint distribution in terms of observational biases.  The observed radius is bounded from below, because very small spots would not cause a detectable signal in the lightcurve.  For values of $y$ further from the transit chord, the smallest detectable radius increases for geometrical reasons: the spot has to overlap with the strip that the planet scans on the stellar surface.  Finally, the spot radius is bounded from above by the physical distribution of spots, and we expect this to be independent from $y$.  However, this is not reflected on Figure \ref{fig:mcmcall}: we see that MCMC states with radius above $0.2$ prefer values of $y$ further from $b$.  We interpret this as an artifact: we suspect that the lightcurve anomaly due to irregularly shaped spots or spot groups, when mistakenly interpreted as a single spot, results in a large spot further from the transit chord.  We believe this radius is unphysical, because we never see transit anomalies that last long enough to require a spot of similar size with $y\approx b$ as an explanation.  Therefore we conclude that the upper limit of physical spot radius distribution is at most somewhat lower than $0.2$, the radius of the largest detected spot with $y\approx b$.  It is also possible that this is not a single spot either, so the actual upper radius limit might be smaller than this.

The \change{four} side panels on Figure \ref{fig:mcmcall} show histograms of spot radius, flux ratio, and \change{$x$ and $y$ coordinates}, generated from the MCMC chains for all spots.  The same observational biases are reflected here: there are no very small ($a\approx0$) spots, and no very bright ($f\approx1$) spots, because these would be undetectable.  \change{$x$ has a bimodal distribution according to the two active latitudes, and} $y$ is concentrated around the impact parameter, because this is where even small spots are eclipsed by the planet.

\begin{figure}
\begin{center}
\includegraphics*[width=80mm]{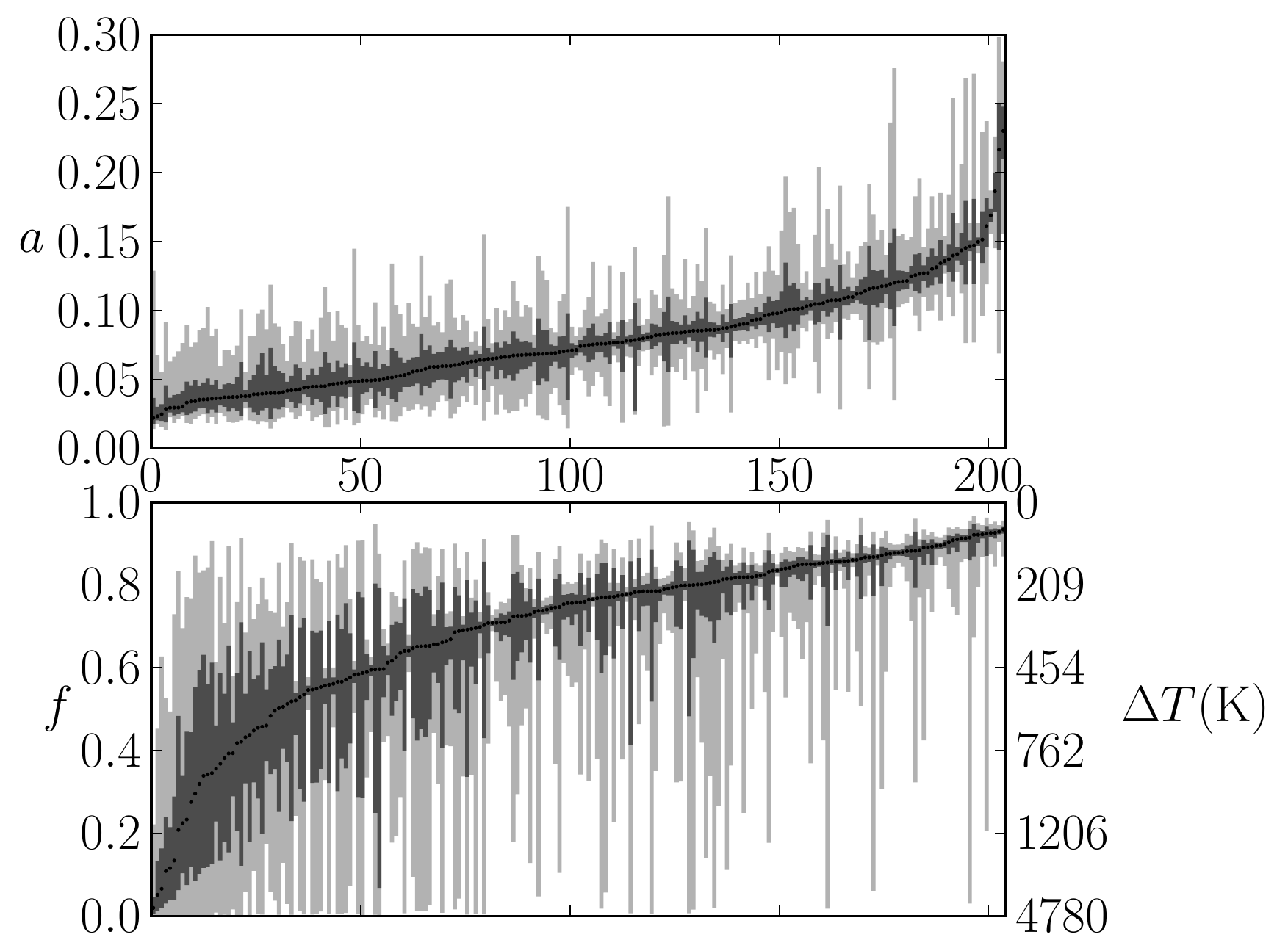}
\end{center}
\caption{Median values of spot radius $a$ (top panel) and flux ratio $f$ (bottom panel).  Horizontal axis shows rank of spot.  Each panel is ordered by the median of the corresponding parameter.  Shaded regions indicate the $1\sigma$ (dark gray) and $3\sigma$ (light gray) confidence intervals based on the MCMC simulation for each spot.  Right vertical axis of bottom panel shows inferred temperature difference of photosphere and spot assuming that both radiate as black bodies.}
\label{fig:afsorted}
\end{figure}

To investigate the radius and flux ratio distribution of spots, we plot the median of these parameters in increasing order on Figure \ref{fig:afsorted}.  The top panel presents spot radius, the bottom shows flux ratio.  The horizontal axes indicate the rank of the spot in the order of median values.  In addition to the median, we shade the $1\sigma$ and $3\sigma$ intervals of the parameter distribution of each individual spot.  The advantage of this presentation over the histograms of Figure \ref{fig:mcmcall} is that we can disentangle the spread of a spot parameter for an individual spot due to uncertainties from the spread of the ensemble distribution due to spots being different.

Together with the histograms of Figure \ref{fig:mcmcall}, Figure \ref{fig:afsorted} helps us confirm the observational biases agains small size and large flux ratio.  In addition, we note that most spots are smaller than $a\approx0.15$ (three times the radius of \hatpelevenb{}\change{, which is approximately $0.059\;R_\star$}), with very few spots around the size of $a\approx0.2$.  We believe that larger spots are artifacts, because they are only seen with $y$ different from $b$.

On the other hand, even though bright spots are more frequent than dark ones, our first impression is that there is a number of almost \change{completely} black spots, that is, spots with $f$ close to zero.  \change{(}Note that a large number density in terms of a spot parameter, that is, large bin count in the histogram corresponds to a less steep curve when data points are plotted in increasing order of that parameter.\change{)}  However, when closely inspecting the parameter uncertainties of individual spots, we see that the ones that seem to be consistent with being very dark show a flux ratio distribution that includes much larger flux ratios as well.  In fact, only two spots have a $3\sigma$ confidence interval that excludes flux ratios above $0.5$, even though 30 spots have median flux ratio and 36 spots have best fit (least sum of squared residuals) flux ratio lower than $0.5$.  \change{On the other hand, more than half of all spots studied have a $3\sigma$ confidence interval that allows flux ratios below 0.5.}  Therefore we cannot either prove or disprove the existence of spots darker than $f=0.5$.  On the other hand, many spots with flux ratios from approximately $0.6$ to $0.9$ have small flux ratio uncertainties, suggesting the existence of spots brighter than $f=0.5$.

\subsection{Spot temperature}

Assuming that the stellar photosphere and the spots both radiate as black bodies, one can calculate the spot temperature $T_\mathrm s$ from the flux ratio $f$ and the effective photospheric temperature $T_\mathrm{eff}$. This method was first applied to infer the spot temperature in the context of spot-induced transit lightcurve anomalies by \citet{2003ApJ...585L.147S}.

We can use the same method to calculate spot temperatures on \hatpeleven{} from \spotrod{} MCMC results.  We use the photospheric effective temperature result $T_\mathrm{eff} = 4780\,\mathrm K \pm 50\,\mathrm K$ of \citet{2010ApJ...710.1724B}, and we integrate Planck's law over the \kepler{} response function to determine the flux ratio as a function of spot temperature.  The temperature difference $\Delta T = T_\mathrm{eff} - T_\mathrm s$ corresponding to certain values of flux ratio $f$ are displayed on the right vertical axis of the bottom panel on Figure \ref{fig:afsorted}.

As we noted in Section \ref{sec:spotparameterdistribution}, we cannot infer anything about the existence of spots with flux ratio less than $0.5$.  Therefore we cannot determine whether there are spots with spot temperature below the corresponding temperature of $T_\mathrm s = 4180\pm40\;\mathrm K$, or a temperature difference of $\Delta T = 600\pm10\;\mathrm K$.

On the other hand, Figure \ref{fig:afsorted} tells us that there are spots with well constrained flux ratios ranging approximately from $f=0.6$ to $0.9$.  In terms of temperature, this means that there exist spots ranging approximately from $T_\mathrm s = 4330\;\mathrm K$ to $4680\;\mathrm K$ ($\Delta T = 450\;\mathrm K$ to $100\;\mathrm K$).  It is possible that \hatpeleven{} also has brighter spots (that are not detected due to observational bias), and darker spots.

For comparison, \citet{2003SoPh..213..301W} show that the Sun exhibits spots ranging from $f\approx0.15$ (with $a\approx0.03$) up to $f\approx0.7$ (with $a\approx0.01$), where the flux ratio is measured at 672.3 nm with 10 nm bandpass.  This shows that \hatpeleven{} is not the only star where individual spots are thought to exhibit a large range of flux ratios.

An advantage of transit anomalies over spectroscopic methods is the ability to measure temperatures of individual spots.  For example, \citet{2004AJ....128.1802O} study TiO absorption features to determine the spot temperature to be $T_\mathrm s = 3350\pm115\;\mathrm K$ on EQ Vir, a BY Dra-type flare star with $T_\mathrm{eff} = 4380\;\mathrm K$, of spectral type K5 Ve, close to K4 of \hatpeleven{}.  This corresponds to a flux ratio of $f=0.21\pm0.05$ in the \kepler{} bandpass.  We note that this value is lower than the flux ratio of the majority of spots we found on \hatpeleven{}, though we were not able to either confirm or exclude the existence of spots with such low flux ratios.  However, the result of \citet{2004AJ....128.1802O} is averaged over all spots visible on the stellar disk, and the temperature range of individual spots cannot be determined by their method.


\subsection{Transit anomaly duration and amplitude}
\label{sec:durationamplitude}

\begin{figure}
\begin{center}
\includegraphics*[width=80mm]{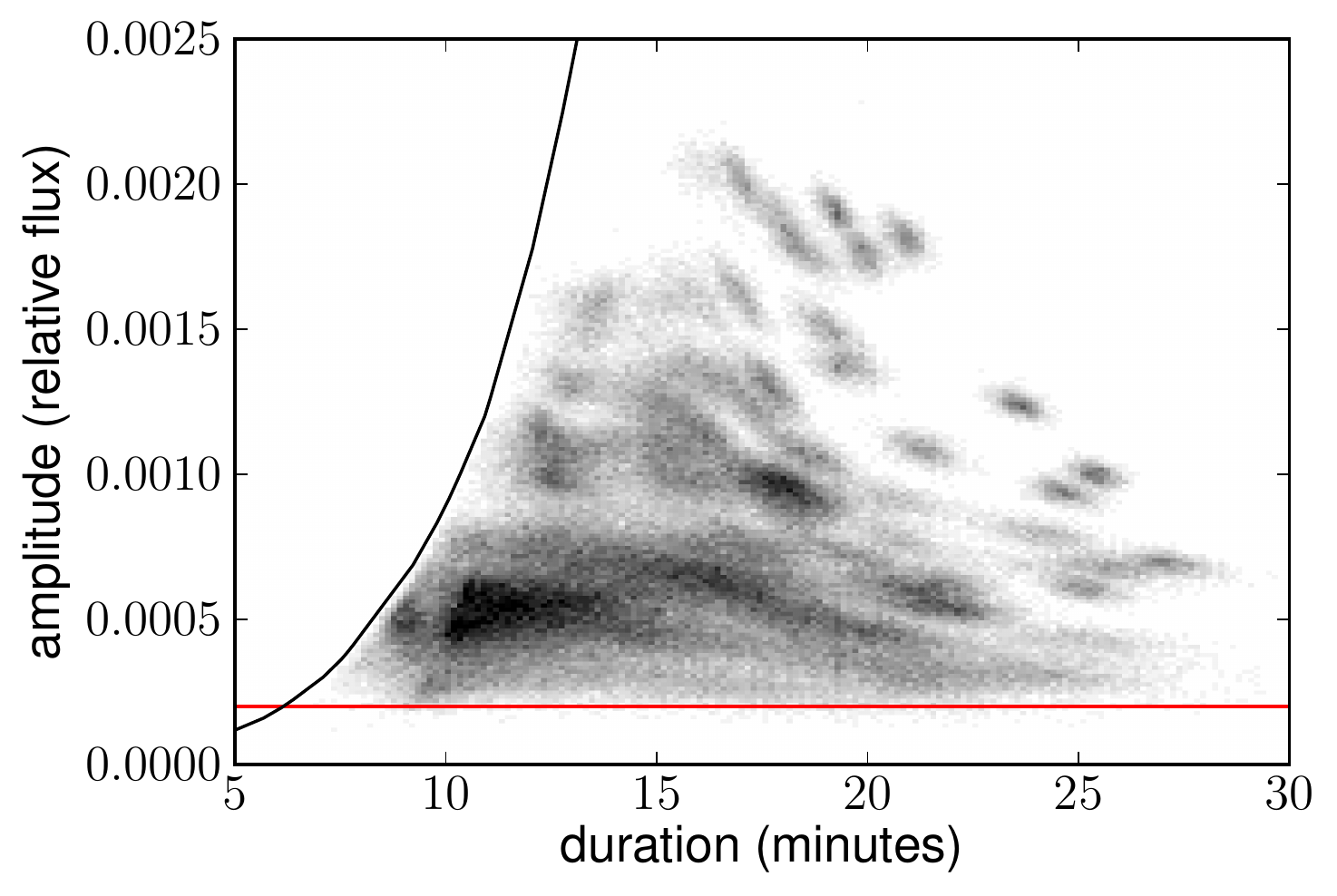}
\end{center}
\caption{Distribution of transit anomaly duration and amplitude in black dots.  Each spot is resampled 1000 times from its MCMC chain.  The $A=0.0002$ constant amplitude line is shown in red.  Black curve represents the boundary of parameter space imposed by our model.}
\label{fig:durationamplitude}
\end{figure}

In addition to investigating the parameters taken by \spotrod{}: $x$, $y$, $a$, and $f$, it is also interesting to study directly observable properties of spot eclipses: the duration and amplitude of the transit anomaly.  Given the results of MCMC calculations, the easiest way to extract these properties is to measure them on model lightcurves.  We therefore draw 1000 states from each chain, and generate model lightcurves with one spot each.  We calculate the amplitude of the transit anomaly as the maximum deviation from a spotless model, and the duration of the transit anomaly as the length of the time interval on which the model with one spot predicts more flux than the spotless model.  The resulting distribution is plotted on Figure \ref{fig:durationamplitude}.

The transit anomaly duration--amplitude distribution is confined from three sides.  The amplitude is bounded from below, as described in Section \ref{sec:spotparameterdistribution}.  We draw a red line on Figure \ref{fig:durationamplitude} at $A=0.0002$, the same amplitude value as on Figure \ref{fig:mcmcall}.  Note, however, that the transit anomaly amplitude is represented directly on Figure \ref{fig:durationamplitude}, whereas on Figure \ref{fig:mcmcall}, we could only calculate the maximum possible value $A_\mathrm{max}$ for given values of $a$ and $f$.

Figure \ref{fig:durationamplitude} tells us that visual inspection as performed by the authors results in an amplitude limit that is mostly constant across anomaly durations.  Anomalies selected programmatically, however, might have a different boundary, as it is possible to identify an anomaly with a smaller amplitude in noisy data if it lasts sufficiently long.

From the side of short anomalies, the distribution is bounded by geometrical arguments.  The amplitude of the anomaly tells us how much flux is missing with respect to a spotless transit.  This places a lower limit on the geometrical extent of the spot, which then translates to the duration of the event.  To characterize this boundary, we feed 10\,000 random black spots with to \spotrod{}, and plot the envelope of their duration-amplitude distribution in black on Figure \ref{fig:durationamplitude}.  Gray spots ($f>0$) would cause anomalies with the same duration but smaller amplitude than if they were black ($f=0$), therefore we only draw black spots to determine this boundary.  We notice that the MCMC distribution extends close to this boundary, which means that the chains contain spot states that are almost entirely black.  As discussed in Section \ref{sec:spotparameterdistribution}, however, this does not mean that the best fit for those spots is necessarily black.

From the right and above, however, the distribution does not extend to its theoretical limits: the maximum conceivable amplitude would be the transit depth (for a black spot that is larger than the planet), and the maximum conceivable transit duration would be the transit duration.  The largest amplitude we observe is roughly half of this, and the longest anomaly lasts only about one quarter of the entire transit.  We interpret these limits as indications of the actual spot parameter distribution, namely the lack of large dark spots, and the upper limit on spot size $a\lessapprox0.2$.  We also note that on Figure \ref{fig:durationamplitude}, the longest transit anomalies seem to have small amplitudes, for which we cannot offer an explanation.

While measuring directly observable quantities like transit anomaly duration and amplitude on models generated by \spotrod{} is very convenient, we note that this inevitably introduces biases.  For example, if we observe an anomaly with duration and amplitude that would place it on the left of the black curve on Figure \ref{fig:durationamplitude}, that could not be reproduced by \spotrod{}, and consequently we would measure different duration and amplitude values with our method.  However, our MCMC analysis disentangles multiple spots and measures transit anomaly durations efficiently, and still yields meaningful conclusions about the darkness of spots and the amplitude threshold for detection.

\subsection{Residuals}

\begin{figure}
\begin{center}
\includegraphics*[width=80mm]{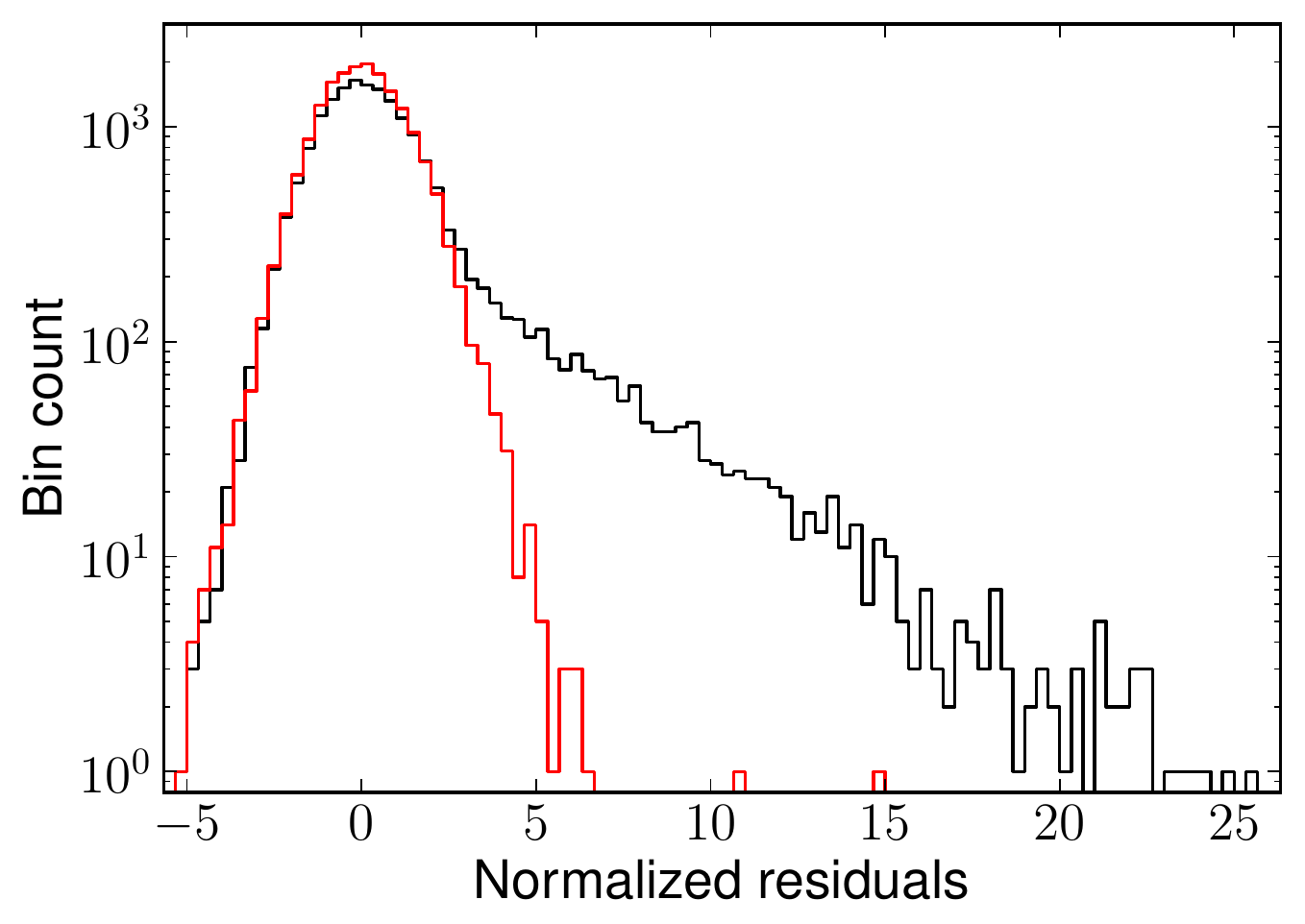}
\end{center}
\caption{Residual histograms of short cadence \kepler{} observations that were taken during transits of \hatpelevenb{}, normalized by \texttt{SAP\_FLUX\_ERR}.  Black histogram corresponds to residuals with respect to the Mandel--Agol lightcurve model, red histogram corresponds to residuals with respect to the best fit \spotrod{} model with a total of \totalspots{} spots.  Vertical axis (counts) is on a logarithmic scale.}
\label{fig:residuals}
\end{figure}

Finally, we study the distribution of residuals to assess the goodness of fit, and to compare our model with \totalspots{} spots to the spotless lightcurve model.  We calculate lightcurves using \spotrod{} and the Mandel--Agol model for the two cases, respectively.  We only consider the residuals at observations that took place during transits (between first and fourth contact).  We use the \texttt{SAP\_FLUX\_ERR} column calculated by the \kepler{} Photometric Analysis module as the error estimate of the lightcurve data in the \texttt{SAP\_FLUX} column.  Figure \ref{fig:residuals} displays the normalized residual distribution histograms: in black when calculated using the spotless model, and in red when calculated using our model with spots, with a logarithmic vertical scale.  The spotless model residuals have a large excess on the positive side, which we attribute to the presence of spot-induced transit lightcurve anomalies.  On the other hand, our \spotrod{} model accounts for enough spots to make the residual distribution fairly symmetric.  We note that this might potentially be used as a detection method to identify targets that exhibit spot-induced transit anomalies.

In terms of the Mandel--Agol fit, it is interesting to note that the largest residual is $25.75\sigma$.  Comparing this to the transit depth of roughly 50 sigma, we see that the largest transit anomaly has an amplitude of half the transit depth, just as we noted in Section \ref{sec:durationamplitude}.  (The exact transit depth expressed in units of photon noise varies due to the quarterly rotations of the \kepler{} satellite.)

The relative error of \hatpeleven{} photometry is very small ($\lessapprox 10^{-4}$).  Assuming that it is dominated by photon noise, the error distribution can be approximated with an independent normal distribution for each data point.  In this case, it is valid to calculate $\chi^2$, and from that, reduced $\chi^2$.  For the models without and with spots, we get the strikingly different values $\chi_\mathrm{spotless}^2=1.7\cdot10^5$ and $\chi_\mathrm{spots}^2=3.3\cdot10^4$, respectively.

The total number of observations taken during the 130 transits is $N=18\,135$.  Since we fix orbital parameters, transit ephemeris, and limb darkening, the spotless model has no fit parameters.  With the number of data points as the degrees of freedom, we get $\chi_\mathrm{red,spotless}^2=9.2$, which, being much larger than unity, motives a model with more free parameters.

Our spot model has four fit parameters for each spot, that is $P=4\cdot\totalspots{}=812$ fit parameters in total.  Note, however, that \citet{2010arXiv1012.3754A} prove that it is not justified to calculate the degrees of freedom as $K=N-P$ for non-linear models like \spotrod{}, and in fact no reliable method is known to calculate $K$ in general.  Therefore we give the value $\frac{\chi_\mathrm{spots}^2}{N-P} = 1.9$ for reference only, and are left with the symmetry of the histogram as the only way to quantify the goodness of the fit.

\section{Conclusion}
\label{sec:conclusion}

In this paper, we present \spotrod{}, a transit lightcurve model accounting for both eclipsed and uneclipsed starspots.  The advantage of our model over previous methods is that in polar coordinates, we integrate analytically with respect to the polar angle, therefore time-consuming numerical integration only remains to be performed along a single dimension, the radial coordinate.  This feature makes our model fast enough not only for fitting, but also for efficient statistical investigations using, for example, MCMC.  A free and open source implementation of our model is publicly available.

The model assumes that spots follow the same limb darkening law as the stellar photosphere, consistent with observations of the Sun \citep{2003SoPh..213..301W}.  It also assumes that spots are homogeneous.  Umbra-penumbra structure can be mimicked by superimposing two concentric spots, while bright features can also be modelled using a flux ratio exceeding unity.

We apply our model to \kepler{} data of \hatpeleven{} transits.  We investigate correlations between fit parameters of individual spots, and confirm findings of previous investigations using similar models.  We also study the size and flux ratio distributions.  We establish an upper limit of $\lessapprox 0.2$ for the spot radius, and find strong indication for the presence of spot with flux ratio ranging from 0.6 to 0.9, corresponding to an effective temperature 100 to 450 K lower than that of the spotless photosphere.  We cannot prove nor disprove the existence of spots with flux ratio less than $0.5$.  We do not find a significant correlation between spot size and flux ratio.

While \hatpeleven{} is unique in its brightness and large transit anomaly amplitudes within the Kepler field, \spotrod{} can potentially be used to model \kepler{} observations of other transiting planetary hosts.  In addition, comparable quality photometric observations are expected to taken of a much larger number of stars, for example, by the K2 mission of the \kepler{} satellite \citep{2014AAS...22322801H}, and by the Transiting Exoplanet Survey Satellite \citep[TESS,][]{2010AAS...21545006R}.

While our model provides a good fit to observations of \hatpeleven{}, it is important to keep in mind that our perfectly circular and homogeneous spots are a simplified version of what the stellar surface actually looks like.  However, mapping out the projected stellar surface by, for example, two dimensional deconvolution with the planetary disk, is an underdetermined and potentially numerically unstable inversion problem.  The general advantage of model simplifications is that the small degree of freedom makes fitting robust.  This is exactly what \spotrod{} provides: a simplistic, approximate, but robust and fast way to model transit lightcurves of spotted stars.

\section*{Acknowledgements}

Work by B.B.~and M.J.H.~was supported by NASA under grant NNX09AB28G from the Kepler Participating Scientist Program and grants NNX09AB33G and NNX13A124G under the Origins program. D.M.K.~is funded by the NASA Carl Sagan Fellowships. This paper includes data collected by the Kepler mission. Funding for the Kepler mission is provided by the NASA Science Mission directorate. The MCMC computations in this paper were run on the Odyssey 2.0 cluster supported by the FAS Science Division Research Computing Group at Harvard University.  B.B.~is grateful to Robert Noyes for useful comments on the manuscript, and to John A.~Johnson for suggesting the name \spotrod{}.

\bigskip

\bibliography{s}

\appendix

\section{Derivations}
\label{sec:derivations}

\subsection{Geometry}
\label{sec:geometry}

\begin{figure}
\begin{center}
\includegraphics*[width=80mm]{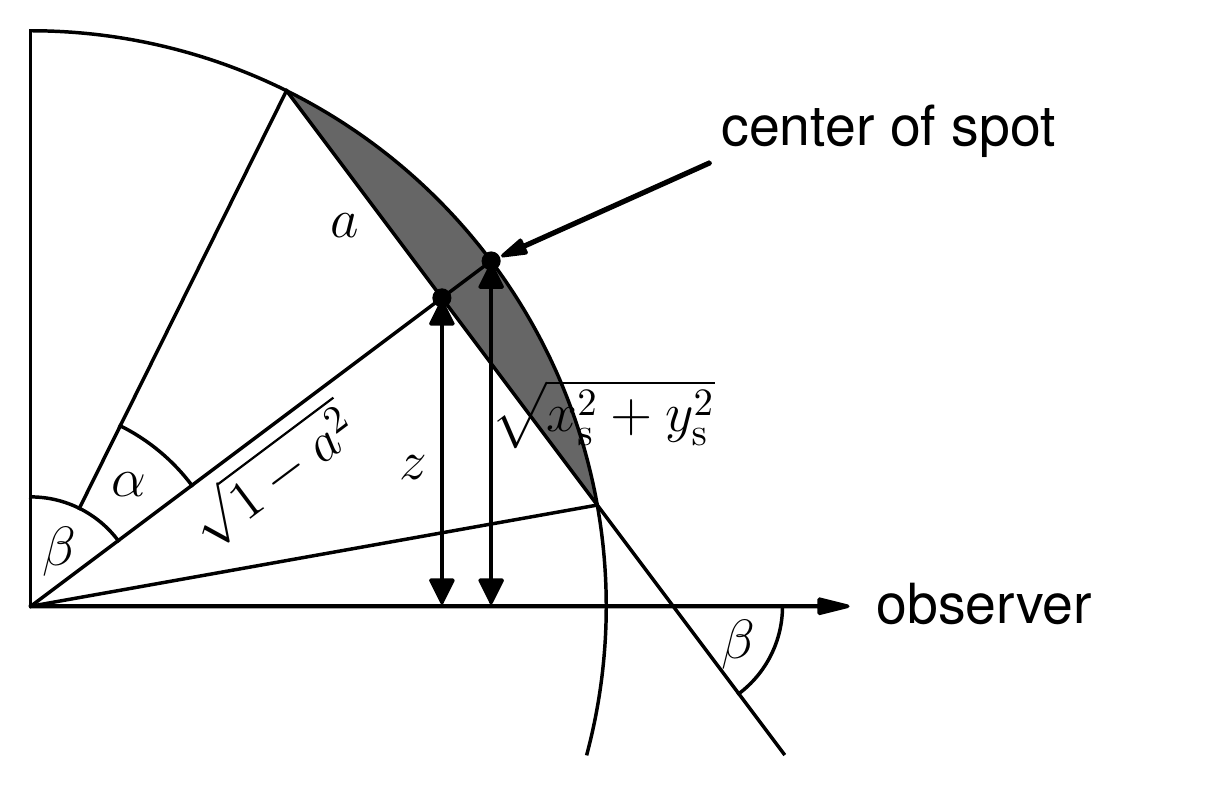}
\end{center}
\caption{A cross section in the plane containing the center of the star, the center of the spot, and the observer. The observer is to the right at an infinite distance.}
\label{fig:sideview}
\end{figure}

Let \xs{} and \ys{} denote the coordinates of the center of the spot as seen by the observer, in a Cartesian coordinate system with the center of the stellar disk as the origin, in stellar radius units. Then the angle between the plane of the spot boundary and the line of sight is 
\begin{align*}
\beta &= \arccos\sqrt{\xs^2+\ys^2}.
\end{align*}
Figure \ref{fig:sideview} shows a side view. Note that this is the same $\beta$ that \citet{2012MNRAS.427.2487K} defines in his Equation (1).

In projection (as seen by the observer), the spot is an ellipse with semi-major axis $a$, and semi-minor axis
\begin{align}
\label{eq:b}
b &= a\sin\beta = a\sqrt{1-(\xs^2+\ys^2)}.
\end{align}
The center of this ellipse is the projection of the intersection point of the axis of the cone and the plane in which the spot boundary lies, not the projection of the center of the spot which is on the stellar surface. The center of the ellipse lies at a distance of $\sqrt{1-a^2}$ from the center of the star, therefore in projection, the distance between the center of the stellar disk and the center of the ellipse is
\begin{align}
\label{eq:z}
z &= \sqrt{1-a^2}\cos\beta = \sqrt{1-a^2}\sqrt{\xs^2+\ys^2}.
\end{align}

Expressing $\xs^2+\ys^2$ from Equation (\ref{eq:z}) and substituting into Equation (\ref{eq:b}), we can express $b$ in terms of $a$ and $z$:
\begin{align}
\label{eq:b1}
b &= a \sqrt{1 - \frac{z^2}{1-a^2}}.
\end{align}

The input parameters of the subroutine \integratetransit{} are \xs{}, \ys{}, and $a$: it calculates $z$ from Equation (\ref{eq:z}), and passes it to \ellipseangle{}, which in turn calculates $b$ using Equation (\ref{eq:b1}).

\subsection{Calculating $\gamma$, $\gamma^*$, and $\delta$}
\label{sec:calc}

\begin{figure}
\begin{center}
\includegraphics*[width=80mm]{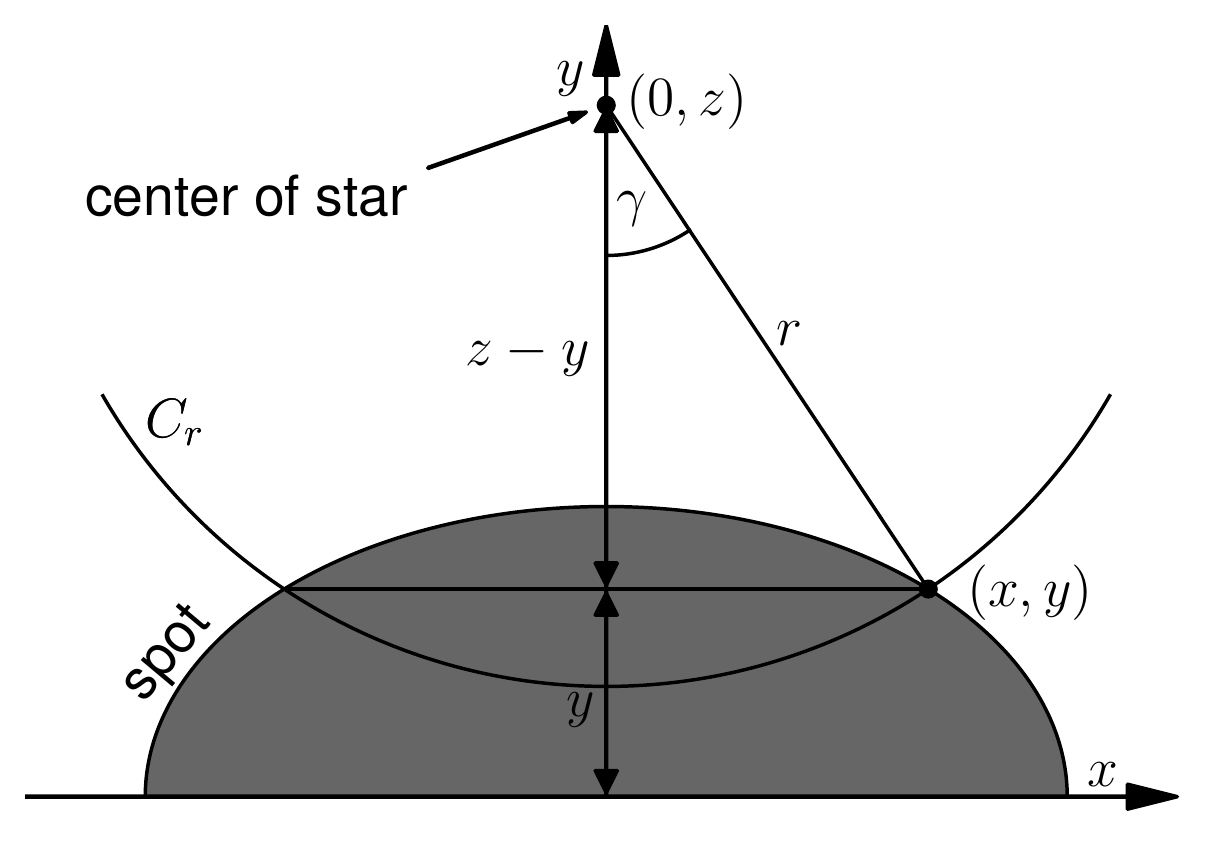}
\end{center}
\caption{The spot, the center of the star, and $C_r$, as seen from the direction of the observer, in a Cartesian coordinate system centered on the center of the spot ellipse (not the projection of the spot center).}
\label{fig:ellipse}
\end{figure}

To calculate $\gamma(r)$, consider a Cartesian coordinate system in the projection plane with the center of the ellipse as the origin, the $x$ axis parallel to the semi-major axis, the $y$ to the semi-minor axis. Let the center of the stellar disk be at $(0,z)$, and let $(x,y)$ denote an intersection point of the ellipse and $C_r$. See Figure \ref{fig:ellipse}. Then $(x,y)$ satisfies the following set of quadratic equations:
\begin{align}
\label{eq:quad1}
\frac{x^2}{a^2} + \frac{y^2}{b^2} &= 1 \\
\label{eq:quad2}
x^2 + (y-z)^2 &= r^2.
\end{align}
If there are no intersection points, then $C_r$ is either located entirely outside the ellipse, thus $\gamma=0$, or entirely \change{inside}, in which case $\gamma=\pi$. If there are one, two, three, or four intersection points, then they must be located symmetrically around the $y$ axis, because if $(x,y)$ is a solution, then so is $(-x,y)$ (and they coincide if $x=0$). After solving for $y$, we calculate $\gamma$ using
\begin{align*}
\gamma &= \arccos\frac{z-y}r.
\end{align*}
Appendix \ref{sec:limb} discusses the case of four intersection points. One or three intersection points are singular cases between other cases, and can be treated along with the case on either side.

\begin{figure}
\begin{center}
\includegraphics*[width=80mm]{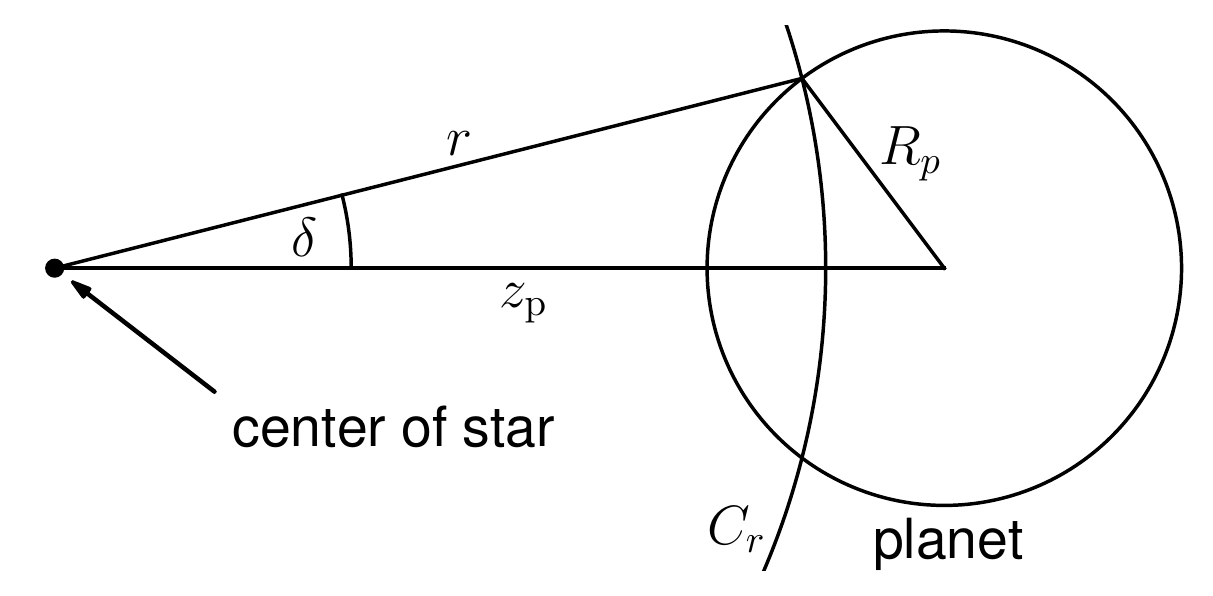}
\end{center}
\caption{The triangle in the sky plane defined by the center of the star, the center of the planet, and one intersection point of the edge of the planet with $C_r$.}
\label{fig:cosines}
\end{figure}

Let $R_\mathrm p$ denote the radius of the planet, and $z_\mathrm p$ its projected separation from the center of the stellar disk, both in stellar radius units. To calculate $\delta(r)$, we could repeat the above derivation with $a=b=R_\mathrm p$. Or we can use the law of cosines in the triangle defined by the center of the star, the center of the planet, and the intersection point of the edge of the planet with $C_r$:
\begin{align*}
R_\mathrm p^2 = r^2 + z_\mathrm p^2 - 2rz_\mathrm p\cos\delta(r),
\end{align*}
as seen on Figure \ref{fig:cosines}. Again, the cases of $C_r$ being disjoint from or entirely occulted by the planet should be tested for separately, yielding $\delta=0$ and $\delta=\pi$, respectively.

\begin{figure}
\begin{center}
\includegraphics*[width=80mm]{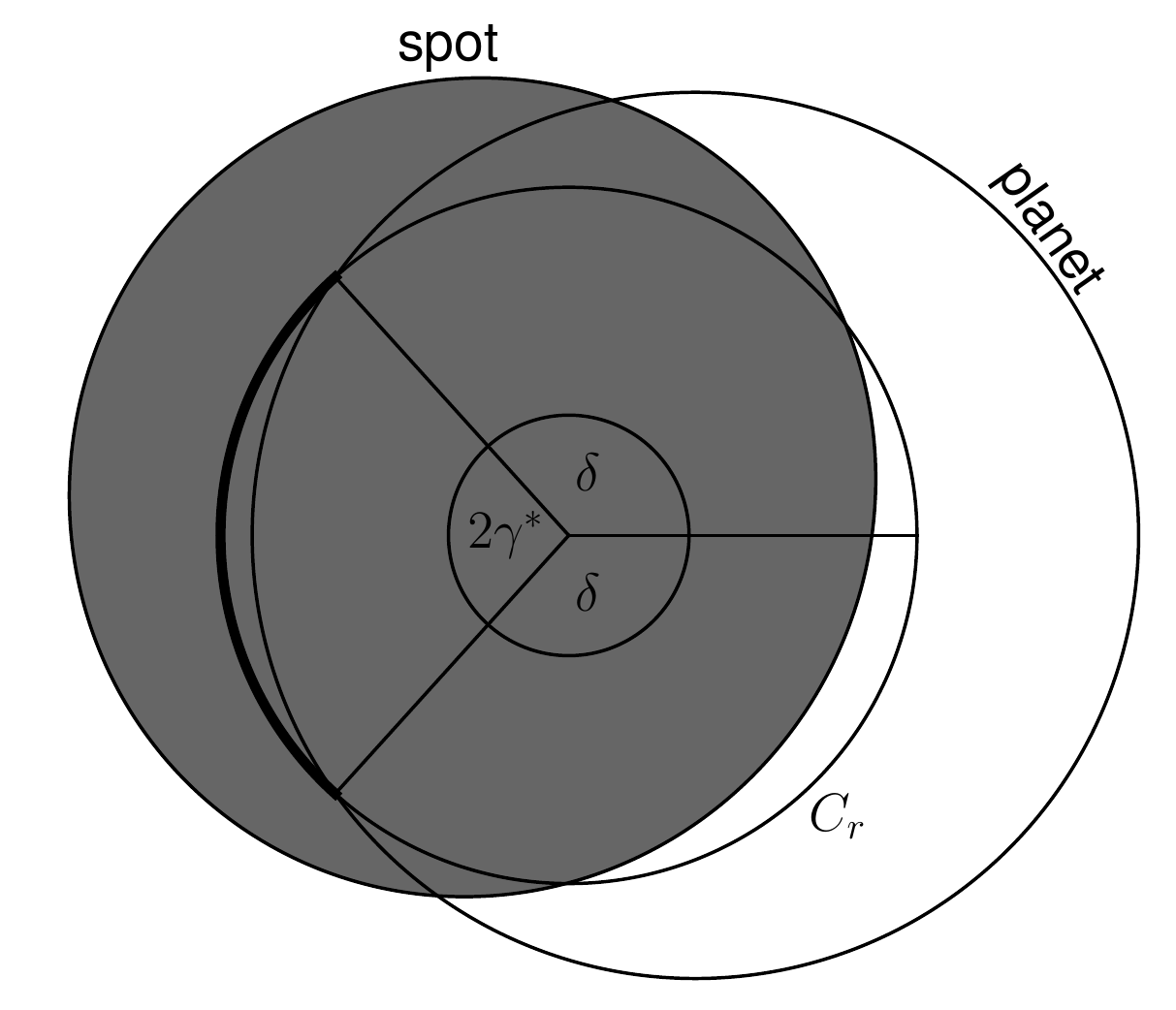}
\end{center}
\caption{Example for the spot arc and the planet arc overlapping in two arcs.}
\label{fig:circular}
\end{figure}

Finally, let $\theta$ denote the angle between the center of the spot and the center of the planet as seen from the center of the star, which we also calculate using the law of cosines. We always choose the angle for which $0\leqslant\theta\leqslant\pi$. Now $\gamma^*(r)$ is determined by $\gamma(r)$, $\delta(r)$, and $\theta$ according to the following cases:
\begin{align}
\label{eq:cases}
\gamma^* = \begin{cases}
\gamma & \textrm{if } \theta \geqslant \gamma + \delta \textrm{ (arcs disjoint)} \\
\gamma - \delta & \textrm{if } \gamma \geqslant \theta + \delta \textrm{ (planet arc inside spot arc)} \\
0 & \textrm{if } \delta \geqslant \gamma + \delta \textrm{ (spot arc inside planet arc)} \\
\frac{\gamma + \theta - \delta}2 & \textrm{o/w, if } \gamma + \delta + \theta \leqslant 2\pi \textrm{ (partial overlap)} \\
\pi - \delta & \textrm{o/w, if } \gamma + \delta + \theta > 2\pi \textrm{ (circular overlap)}.
\end{cases}
\end{align}
In the first case, the arcs are disjoint, therefore none of the spot arc is eclipsed by the planet. This happens to spot 1 on Figure \ref{fig:observer}. 
In the second case, the entire planet arc gets subtracted from the spot arc. The part of the spot arc that is not occulted by the planet is now composed of two arcs on either side of the planet, and we define $\gamma^*$ as half the total central angle of them. In the third case, the planet occults the entire spot arc. ``Otherwise'' for the last two cases means that the triangle inequality holds between $\gamma$, $\delta$, and $\theta$. In the fourth case, the planet and spot arcs overlap in a single arc. This happens to spot 2 on Figure \ref{fig:observer}. In the last case, they overlap in two arcs. This can happen only if at least one of the planetary disk and the spot ellipse contain the center of the stellar disk in projection. This situation is illustrated by Figure \ref{fig:circular}.

The function \circleangle{} calculates $\beta$ from $r$, $z_\mathrm p$, and $R_\mathrm p$, \ellipseangle{} calculates $\gamma$ from $r$, $z$, and $a$, and finally \integratetransit{} calculates $\gamma^*$ based on Equation (\ref{eq:cases}), evaluates the integrals of Equations (\ref{eq:multispot}) and (\ref{eq:transit}), and calculates $F_\mathrm{normalized}$ according to Equation (\ref{eq:normalized}).

\subsection{Spots partially behind the limb}
\label{sec:limb}

The boundary of the spot is assumed to be a circle, therefore its projection (as seen by the observer) is an ellipse. However, we must investigate whether the boundary of what we see of the spot coincides with the projection of its boundary on the stellar surface. It is easy to see that it is indeed the case as long as no part of the spot covers another part of it. That is, as long as the entire spot is visible, we will always see it as an ellipse. (Remember the caveat that the center of this ellipse is not the projection of what we defined as the center of the spot.)

\begin{figure}
\begin{center}
\includegraphics*[width=80mm]{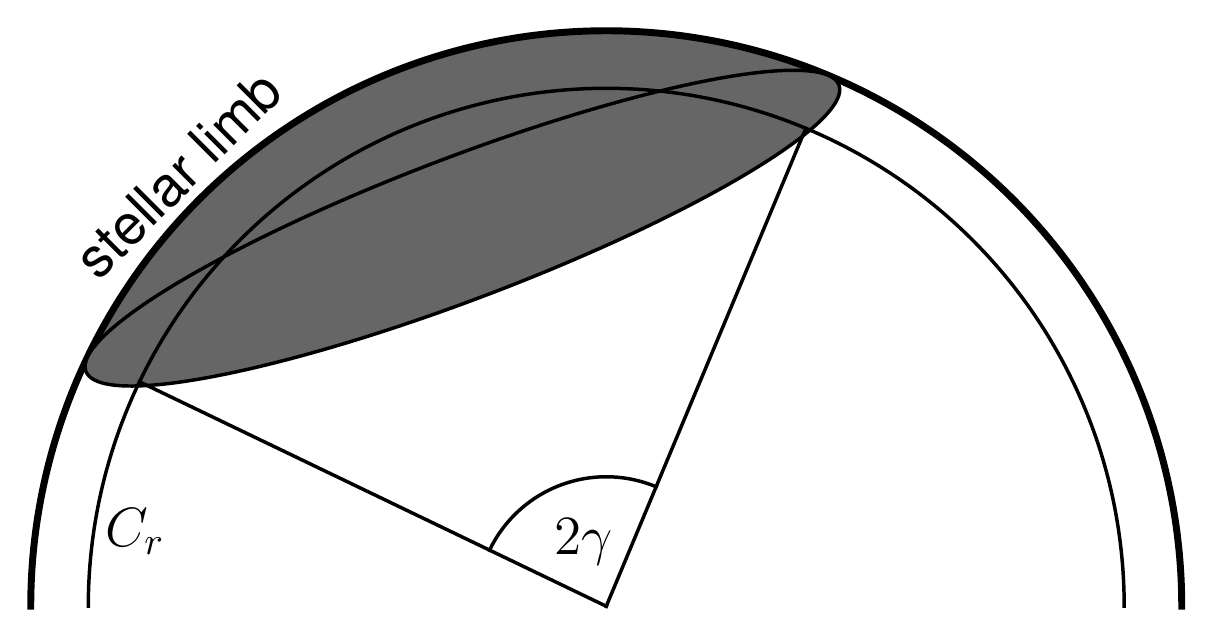}
\end{center}
\caption{Example of a spot partially hiding behind the stellar limb. Note how it is bounded partially by an arc of the ellipse that is its boundary in projection, and partially by an arc of the stellar limb.}
\label{fig:limbspot}
\end{figure}

However, when the spot partially hides behind the stellar limb, we will see it as a shape bounded by an arc of an ellipse (the projection of the spot boundary) and an arc of a circle (the edge of the stellar disk), as illustrated on Figure \ref{fig:limbspot}. We prove in the next section that in this case, the ellipse touches the stellar limb from the inside.

The two cases (the entire spot visible, or part of it is behind the limb) are delineated by a critical value of $\beta$, or equivalently, a critical value of $z$, both of which we expect to depend only on $a$. From Figure \ref{fig:sideview}, we can see that in this critical case, 
\begin{align*}
\beta_\mathrm{crit} &= \alpha \\
\zcrit &= \sqrt{1-a^2} \cos\beta_\mathrm{crit} = \sqrt{1-a^2} \cos\alpha \\
&= \sqrt{1-a^2} \cos\arcsin a = \sqrt{1-a^2} \sqrt{1-a^2} \\
&= 1-a^2,
\end{align*}
using Equation (\ref{eq:z}) to express $z$ in terms of $a$ and $\beta$.

If $z<\zcrit$, then the entire spot is visible. If $z=\zcrit$, then the ellipse touches the edge of the stellar limb at one point, as we prove in the next section. If $z>\zcrit$, then part of the spot is behind the stellar limb. Note that since we describe the spot with the parameter $z$ (implicity, through $x_s$ and $y_s$), $z=1$ corresponding to the spot center being on the stellar limb, therefore we cannot deal with spots that are partially visible, but their center is behind the limb. Such spots, however, will have a small contribution to $F_s$ and $F_\mathrm{transit}$, because only a very small part of them is visible, furthermore this small part is on the limb, which is usually darker to start with.  Furthermore, since $F_s\approx F_\mathrm{transit}$ as long as $R_\mathrm p\ll1$, omitting such a spot will have a very small effect on $F_\mathrm{normalized}$.

\subsection{Number of intersection points}
\label{sec:number}

In this section, we prove that as long as $z<\zcrit$, the spot boundary and $C_r$ can have at most two intersection points. We also explain how to correctly model the ellipse in terms of $\gamma$ if $z>\zcrit$ and there are four intersection points. Finally, we prove that if $z=\zcrit$, then the spot boundary ellipse touches the stellar limb at one point, and if $z>\zcrit$, then at two points. 

Let us express $x^2$ from Equation (\ref{eq:quad1}) and substitute into Equation (\ref{eq:quad2}). This results in a quadratic equation for $y$:
\begin{align}
\label{eq:quadratic}
0 &= A y^2 + B y + C \\
\nonumber
A &= \frac{a^2}{b^2} - 1 \\
\nonumber
B &= 2z \\
\nonumber
C &= r^2 - a^2 - z^2 \\
\nonumber
y_\pm &= \frac{-z \pm \sqrt{z^2 - \left(\frac{a^2}{b^2}-1\right)\left(r^2-a^2-z^2\right)}}{\frac{a^2}{b^2}-1}.
\end{align}
Equation (\ref{eq:quad1}) tells us that each solution $y$ represents two intersection points if $|y|<b$, one if $|y|=b$, or none if $|y|>b$.
Now let us consider the following inequality:
\begin{align}
\label{ineq:proveme}
y_- &< -b \\
\frac{-z - \sqrt{z^2 - \left(\frac{a^2}{b^2}-1\right)\left(r^2-a^2-z^2\right)}}{\frac{a^2}{b^2}-1} &< -b.
\end{align}
This inequality is a sufficient condition for that $y_-$ does not represent real intersection points, that is, there are at most two intersection points (corresponding to $y_+$) 
Now we increase the left hand side of Inequality (\ref{ineq:proveme}). This will make it sharper, leading to a more restrictive, therefore still sufficient (but not necessary) condition. If we find at the end that this still holds whenever $z<\zcrit$, that proves our original statement: 
\begin{align}
\label{ineq:root}
\frac{-z}{\frac{a^2}{b^2}-1} &< -b.
\end{align}
Assume that $z>0$, in which case we also have $a>b$ and thus $A>0$. Then multiplying by $\frac{Ab}{az}$ does not change the direction of inequality:
\begin{align}
\nonumber
-\frac ba &< -\frac{b^2}{az} \left(\frac{a^2}{b^2}-1\right) \\
\label{ineq:samecrit}
\frac az - \frac{b^2}{az} &< \frac ba.
\end{align}
Substituting $b$ from Equation (\ref{eq:b1}), we get:
\begin{align}
\nonumber
\frac az - \frac az \left(1 - \frac{z^2}{1-a^2}\right) &< \sqrt{1 - \frac{z^2}{1-a^2}} \\
\label{ineq:gettingthere}
\frac{az}{1-a^2} &< \sqrt{1 - \frac{z^2}{1-a^2}}.
\end{align}
Both sides of Inequality (\ref{ineq:gettingthere}) are positive, therefore we can square them to get an equivalent inequality:
\begin{align}
\nonumber
\frac{a^2z^2}{\left(1-a^2\right)^2} &< 1 - \frac{z^2}{1-a^2} \\
\nonumber
a^2z^2 &< (1 - a^2)^2 - (1-a^2) z^2 \\
\nonumber
z^2 &< \left(1-a^2\right)^2 \\
\nonumber
z &< 1-a^2 \\
\label{ineq:final}
z &< \zcrit.
\end{align}
We are also allowed to multiply by $(1-a^2)^2$, which has to be positive. Finally, we arrive exactly at the critical value of $z$ that we have already established.

This derivation shows that if Inequality (\ref{ineq:final}) holds, then so does Inequality (\ref{ineq:proveme}). That is, if the spot is entirely visible, then there are at most two intersection points.

\begin{figure}
\begin{center}
\includegraphics*[width=80mm]{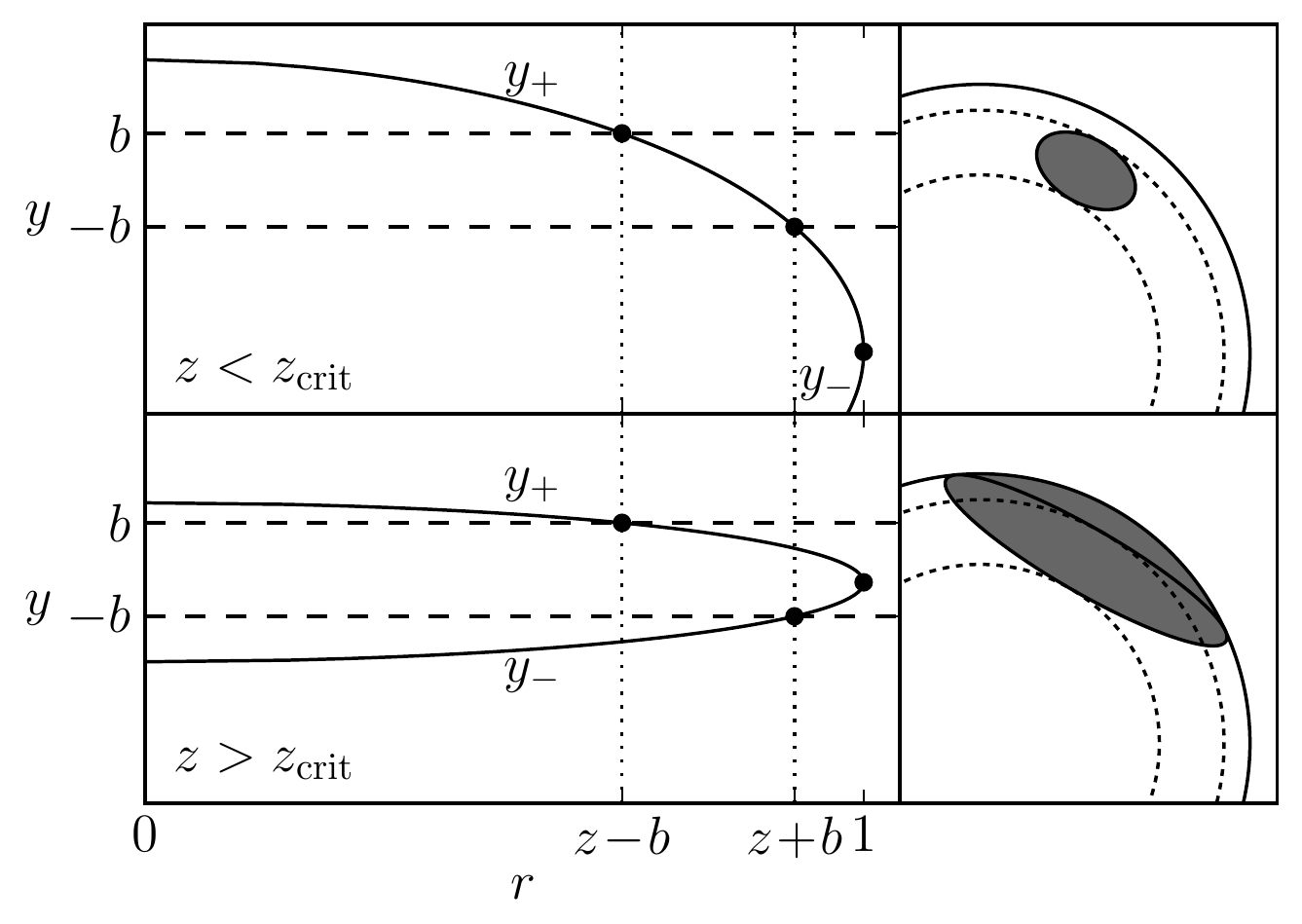}
\end{center}
\caption{Left panels: $y_\pm$ as a function of $r$. The upper branch of the parabola is $y_+$, the lower branch is $y_-$. Horizontal dashed lines are drawn at $y=\pm b$, vertical dotted lines at $r=z\pm b$ where $|y|=b$. Right panels: spots in projection, with $C_{r_1}$ and $C_{r_2}$, the two circles that touch the projection of the boundary of the spots (not the boundary of their projection), in dotted lines. Top panels: $a=0.2$, $\zcrit=0.96$, the spot is entirely visible. Bottom panels: $a=0.6$, $\zcrit=0.64$, the spot is partially behind the limb. In both cases, $b=0.12, z=0.78, r_1=0.66, r_2=0.90$.}
\label{fig:ypm}
\end{figure}

To understand the dependence of the number of intersection points on $r$, we plot $y_\pm$ as a function of $r$ on Figure \ref{fig:ypm}. So far we have proven that if $z<\zcrit$, then $y_-<-b$ for all values of $r$, which case is illustrated on the top panels. We have one intersection point if and only if $|y_+|=b$, which happens at $r=z\pm b$, when $C_r$ touches the ellipse. 

In case $z>\zcrit$, let us investigate how to properly account for the spot shape. If $r<z-b$, then $C_r$ is disjoint from the spot. At $r=z-b$, the circle $C_r$ touches the ellipse, we have $y_+=b$, yielding $x=0$ as a multiple root for the intersection point. Further increasing $r$ will result in $y_+<b$, representing two distinct real solutions for $x$. When $r$ reaches $z+b$, $C_r$ will touch the ellipse from the inside at a point corresponding to $y_-=-b$, $x=0$: this is the third point of intersection.

This can happen if and only if the radius $R$ of the osculating circle at the endpoint of the semi-minor axis is larger than $r=z+b$. As a sanity check, let us investigate what it means in terms of $z$:
\begin{align*}
R &> r \\
\frac{a^2}b &> z + b.
\end{align*}
This inequality is $\frac{az}b$ times Inequality (\ref{ineq:samecrit}) with the inequality sign in the other direction. Inequality (\ref{ineq:samecrit}), in turn, is equivalent to Inequality (\ref{ineq:final}). That is, the condition on the curvature of the ellipse is equivalent to Inequality (\ref{ineq:final}) with the inequality sign in the other direction: $z>\zcrit$. This is consistent with our previous statements.


Further increasing $r$ will yield four distinct intersection points, with $-b<y_-<y_+<b$. This scenario is also demonstrated on Figure \ref{fig:limbspot}. By continuity, the outside pair of intersection points corresponds to $y_+$, because they exist ever since $r=z-b$. The inside pair only appeared at $r=z+b$, and therefore corresponds to $y_-$, which crossed $-b$ at the same value of $r$.

Our code always calculates $\gamma$ based on $y_+$. This means that the entire arc between the outside intersection points is considered to be part of the spot, which correctly describes the shape of the spot as the observer sees it. That is, the code gives the correct result even in case $z>\zcrit$.

Finally, if $r=1$, then the inside and outside pair of intersection points coincide according to Figures \ref{fig:limbspot} and \ref{fig:ypm}: the ellipse touches the stellar limb from the inside. Another way of saying this is $y_-=y_+$, which happens exactly if the discriminant of Equation (\ref{eq:quadratic}) is zero. We now prove this statement.
\begin{align*}
B^2 - 4AC &= 0 \\
4z^2 - 4 \left(\frac{a^2}{b^2} - 1\right) \left(r^2 - a^2 - z^2\right) &= 0 \\
z^2 - \left(\frac1{1 - \frac{z^2}{1-a^2}} - 1\right) \left(1 - a^2 - z^2\right) &= 0 \\
z^2 - \left(\frac{1 - a^2}{1 - a^2 - z^2} - 1\right) \left(1 - a^2 - z^2\right) &= 0 \\
z^2 - \frac{z^2}{1 - a^2 - z^2} \left(1 - a^2 - z^2\right) &= 0.
\end{align*}
The last equation trivially holds true, which proves that if $r=1$, then $y_- = y_+$. Note that this is true regardless of the value of $z$: if $z<\zcrit$, then $y_- = y_+ < -b$ (no real solution for $x$, no intersection points); if $z=\zcrit$, then $y_- = y_+ = -b$ (touching in a single point with $x=0$, quadruple root); and if \change{$z>\zcrit$}, then $y_- = y_+ > -b$ (touching at two points, multiple roots each, like on Figures \ref{fig:limbspot} and \ref{fig:ypm}), as seen from Ineqalities (\ref{ineq:root}--\ref{ineq:final}). This concludes our proof.

\bsp

\label{lastpage}

\end{document}